\documentclass[a4paper,superscriptaddress,amsmath,amssymb,11pt]{article}

\usepackage{jcappub} 

\usepackage[T1]{fontenc} 
\usepackage[utf8]{inputenc}
\usepackage{graphicx}
\usepackage{dcolumn}
\usepackage{bm}
\usepackage{float}
\usepackage{multirow}
\usepackage{verbatim}
\usepackage{amsmath}
\usepackage{graphicx,subfigure,epsfig}
\usepackage{mathrsfs}
\usepackage{array}
\usepackage{caption}   
\usepackage{hyperref}
\hypersetup{
    colorlinks=true,
    citecolor=blue,
    linkcolor=blue,
    filecolor=magenta,
    urlcolor=black,}
\usepackage{color}

\newcommand{\be}{\begin{equation}}
\newcommand{\ee}{\end{equation}}

\title{Probing the black holes in a dark matter halo of M87 using gravitational wave echoes} 

\author[a]{Dong Liu,}
\author[b]{Yi Yang,}
\author[a,1]{Zheng-Wen Long\note{Corresponding author.}}

\affiliation[a]{College of Physics, Guizhou University, Guiyang, 550025, China}
\affiliation[b]{School of Mathematics and Statistics, Guizhou University of Finance and Economics, Guiyang, 550025, China}

\emailAdd{dongliuvv@yeah.net}
\emailAdd{yiyang@mail.gufe.edu.cn}
\emailAdd{zwlong@gzu.edu.cn}

\abstract{Variations at the event horizon structure of a black hole will emit the signals of the gravitational wave echoes associated with the ringdown of the binary black hole merger. In this work, combining mass model of M87 and Einasto profile for dark matter halo, we construct one formal solution of black holes in dark matter halo, and this solution includes the regular black hole for geometric parameter $a < 2M$ and the wormhole for $a \geq 2M$. Then, we test this solution under axial gravitational perturbation and calculate their quasinormal modes (QNM). Our results show that when geometric parameter $a>2M$, a series of gravitational wave echoes appear after the QNM, and one distinctive feature of gravitational wave echoes is the double barrier. Besides, we also study the impacts of shape parameters $\alpha$ of Einasto profile both on the QNM and gravitational wave echoes, and extract their frequencies using WKB method and Prony method. In principle, with the increasing of shape parameter $\alpha$, the frequencies of the QNM and gravitational wave echoes between the different shape parameters must be different. But now there is zero difference of the frequencies between shape parameter $\alpha=0.18$ and $\alpha=0.20$. Zero difference is not allowed, so we give an upper limit for shape parameter $\alpha$, which is approximately $\alpha<0.18$. These signals of gravitational wave echoes in a dark matter halo may be detected in the near future, and these characteristics of gravitational wave echoes may serve as local measurements of dark matter.}


\begin{document}
\maketitle
\flushbottom

\section{Introduction}\label{s1}
The black hole is a beautiful and mysterious celestial object predicted by Einstein in General Relativity (GR). The gravitational field of such a black hole is so powerful that nothing within the event horizon, not even light, can escape.  Since 2016, LIGO/Virgo Collaboration have detected a series of gravitational wave events emitted from the merger of a binary black hole in our universe \cite{LIGO2, LIGO1, LIGO3, LIGO4}. Besides, the Event Horizon Telescope (EHT) Collaboration released the first images of a black hole in the M87 and Sgr A, respectively \cite{EventHorizonTelescope:2019dse, EventHorizonTelescope:2022apq}. Based on these observational evidences, the existence of black holes in the universe is generally accepted. For a merger event of binary black holes in gravitational wave events, it usually goes through three stages: Inspiral, Merger and Ringdown \cite{LIGOScientific:2016aoc}. Among them, the ringdown stage can be described by a superposition of complex frequency damping exponents, which is known as quasinormal mode (QNM). Generally speaking, the process of QNM typically corresponds to a dissipative system in physics, which is accompanied by energy loss. For black holes, it can describe the process of energy radiation through gravitational waves following a perturbation. Therefore, the QNM can be regarded as a special characteristic of black holes and it can be used as a tool to test GR and identify black holes in our universe \cite{Konoplya:2011qq,Rahman:2021kwb}. In other words, the QNM is also considered a special fingerprint of black holes. Based on these special fingerprints, LIGO/Virgo Collaboration may be able to detect some new physical properties of special black holes, such as the gravitational wave echoes. The appearance of gravitational wave echoes is usually closely related to the horizon structure of exotic compact objects, such as wormholes \cite{Simpson:2018tsi, Simpson:2019cer, Churilova:2019cyt,Pal:2022cxb}, gravastars \cite{Mazur:2001fv} and firewalls \cite{Almheiri:2012rt}. The time evolution of gravitational wave echoes usually appears itself as a series of echoes after the QNM. Therefore, as an observable phenomenon, gravitational wave echoes can help ones better understand the true horizon structure of the black hole \cite{Vicente:2018mxl, Bueno:2017hyj, Maselli:2017cmm, Cardoso:2016oxy, Cardoso:2016rao,Dong:2020odp,Yang:2021cvh}. Currently, the main methods used to calculate the QNM of a black hole include the time-domain method \cite{PhysRevD.49.883}, the WKB method \cite{Schutz:1985km,PhysRevD.35.3621}, the P{\"o}schl-Teller potential approximation \cite{Poschl:1933zz,Ferrari:1984zz,Churilova:2021nnc} and the continued fraction method \cite{Leaver:1985ax,Dolan:2007mj}. Among them, the WKB method offers an analytical approximation approach to determine the QNM frequencies through asymptotic expansion. Based on this, G. Panotopoulos and $\acute{\text{A}}$. Rinc$\acute{\text{o}}$n \cite{Panotopoulos:2017hns,Rincon:2018sgd,Panotopoulos:2019qjk} studied the QNM of the black holes in Einstein-power-Maxwell theory and regular black hole using WKB method. Ramin G. Daghigh and Michael D. Green \cite{Daghigh:2011ty} studied  the validity of the WKB Approximation method in the QNM of the non-rotating black hole spacetimes. J.A. V. Campos et al. \cite{Campos:2021sff} studied the  QNM of the noncommutative black hole. J. Matyjasek and M. Telecka \cite{Matyjasek:2019eeu} proposed an improvement of the WKB metnod to study the QNM of the black holes.


On the other hand, the natural properties of dark matter is one of the most profound questions in astrophysics and cosmology. So far, there is a large number of studies that indirectly indicate the existence of dark matter, such as the cosmic microwave background radiation, the rotation curve of galaxies, and the large-scale structure of the universe. The distribution of dark matter in the large-scale structure of galaxies is clear, but its distribution near the galaxy cores and supermassive black holes is unclear. F. Navarro et al. \cite{Navarro:1995iw} first used N-body simulations to study the distribution of the cold dark matter halo, that is NFW profile. However, W.J.G. de Blok et al. \cite{deBlok:2009sp} discovered through observations and cosmological simulations that the NFW density profile has a ``cusp'' problem at small-scales structure. Different from the structure of ``cusp'', P. Gondolo et al. \cite{Gondolo:1999ef} have shown that due to the strong gravity of the black hole itself, the cold dark matter in the center of the Milky Way will be accreted by the black hole into a ``spike'' structure. However, there is currently no clear evidence for the existence of ``cusp'' and ``spike'' in the center of the Milky Way. Therefore, studying the distribution of dark matter near black holes is an important and interesting topic. Regarding some new researches on the dark matter halo and black holes, we list these below. Z. Xu and D. Liu et al. \cite{Xu:2017bpz,Xu:2017vse,Xu_2018,Xu:2021dkv,Liu:2023oab,Liu:2022ygf,Yang:2023tip,Liu:2024xcd} constructed the black hole metric in the dark matter halo based on different dark matter density profiles, and extended them to the case of rotation. C. Zhang et al. \cite{Zhang:2021bdr, Zhang:2022roh} studied the QNM of Schwarzschild-like black holes in different dark matter haloes of M87, and also studied the impacts of dark matter parameters on QNM. K. Jusufi et al. \cite{Jusufi:2019nrn,Vagnozzi:2022moj} studied the black hole shadow of a Kerr-like black hole in a dark matter halo of M87 and studied the impacts of dark matter parameters on the black hole shadow. Based on fully-relativistic formalism, V. Cardoso et al. \cite{Cardoso:2022whc} gave a general solution to the gravitational perturbation of black holes in astrophysical environments, which includes an accretion disk or a dark matter halo. They then applied this method to the dark matter environment \cite{Duque:2023cac,Speeney:2024mas} and calculated the quasinormal mode of black holes under the gravitational perturbation in a dark matter halo \cite{Cardoso:2021wlq}. After that, based on the exact solution of black holes in Ref. \cite{Cardoso:2021wlq}, R. Konoplya et al.  \cite{Konoplya:2021ube} calculated the quasinormal modes of black holes, the greybody factors and the Unruh temperature under the electromagnetic field perturbation. Then based on the fully-relativistic formalism, R. Konoplya et al. \cite{Konoplya:2022hbl} obtained an exact solution to the black hole metric from various density profiles of dark matter halo. Z. Shen et al. \cite{Shen:2023erj} gave five black hole models immersed in a dark matter haloe, and extended these results to the case of rotating black hole. E. Figueiredo et al. \cite{Figueiredo:2023gas} tested the impacts of gravitational waves on different dark matter density profile parameters through the axial gravitational perturbation. V. Xavier et al. \cite{Xavier:2023exm} used the data of the black hole shadow in the M87 and Sgr A to study and constrain the parameters of a dark matter halo. The research results of these authors all play a positive role in the indirect detection of dark matter from the perspective of the black hole.

If there is dark matter near the black hole, it may significantly alter the spacetime characteristics around the black hole, leading to deviations from the black hole in GR. Therefore, the prospect of using the QNM of a black hole to test these deviations is particularly interesting. It can help us to more fully understand the nature of black holes and their surroundings, and thereby explore possible new physical phenomena. In point of fact, gravitational wave echoes, as one of the observable phenomenons, may be used for studying and detecting black holes in the universe. Gravitational wave echoes typically appear as a series of echo signals following the QNM stage \cite{Churilova:2019cyt}. Therefore, in this work, combining the mass model of M87 and the Einasto profile for a dark matter halo, we construct one formal solution both for the regular black hole and the wormhole in a dark matter halo. For some relevant research on regular black holes without dark matter, one can refer to Refs. \cite{Hayward:2005gi,Ayon-Beato:1999kuh,Abdujabbarov:2016hnw,Yang:2023agi,Balart:2023odm,Konoplya:2024hfg}. Then, studying the impacts of dark matter on the QNM. Meanwhile, planning to investigate how dark matter affects the signal of the gravitational wave echoes. Finally, studying and discussing whether these impacts of dark matter on a black hole has the possibility of detection. To a certain extent, with the help of black holes to study dark matter may provide more possibilities for local measurements of dark matter. Finally, we hope that these new findings on QNM and gravitational wave echoes can provide a new thought for indirect detection of dark matter.

The organization of this paper is as follows. In Section \ref{s2}, combining the mass model of M87 and the Einasto profile of dark matter, we give the metric of pure dark matter spacetime and compare it with Minkowshi spacetime. In Section \ref{s3}, starting from Einstein field equations, we constructed one formal solution both for a regular black hole and a traversable wormhole in a dark matter halo. In Section \ref{s4}, we test this solution under the axial gravitational perturbation and preliminarily study the impacts of the geometric parameters of black holes on the effective potential. In Section \ref{s5}, we introduce the numerical methods used to calculate QNM and gravitational wave echoes, and show the relevant results of QNM and gravitational wave echoes of black holes in dark matter haloes. Finally, we study and discuss the possibility of detecting the impact of dark matter on QNM under the astronomical observation. Section \ref{s6} is our summary and conclusion. 

The halo density and characteristic radius in the M87 are $\rho_e=6.9 \times 10^6 M_\odot/kpc^3, r_e=91.2kpc$ and the mass of the black hole at the center of M87 is $M_\text{BH}=6.5 \times 10^9 M_\odot$ \cite{EventHorizonTelescope:2019dse}. In this paper, we use the natural units of $G=c=1$ and black hole units. All the Greek indices run from 1 to 4.

\section{The spacetime of the pure dark matter fitting by the Einasto profile}\label{s2}
\begin{figure*}[t!]
\centering
{
\label{fig:b}
\includegraphics[width=0.3\columnwidth]{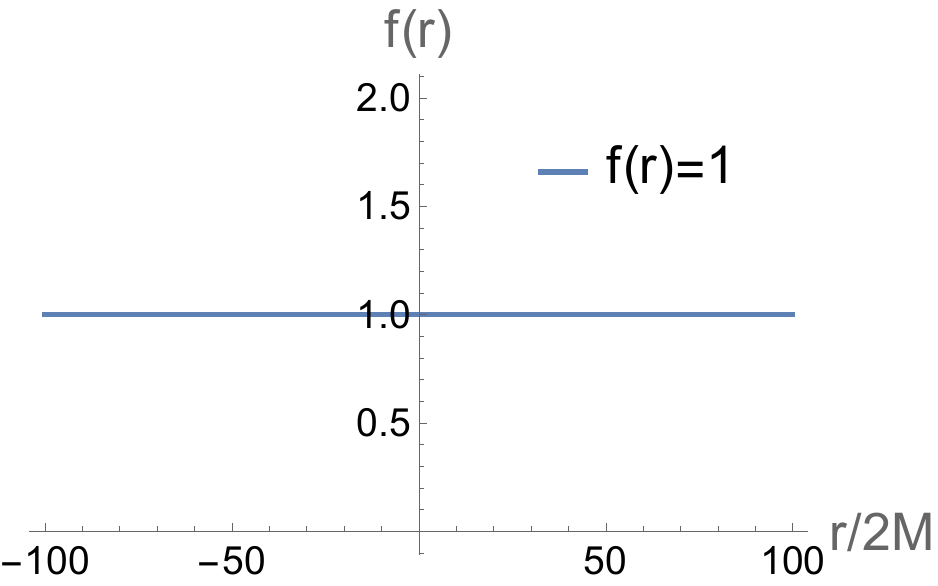}
}
{
\label{fig:b}
\includegraphics[width=0.3\columnwidth]{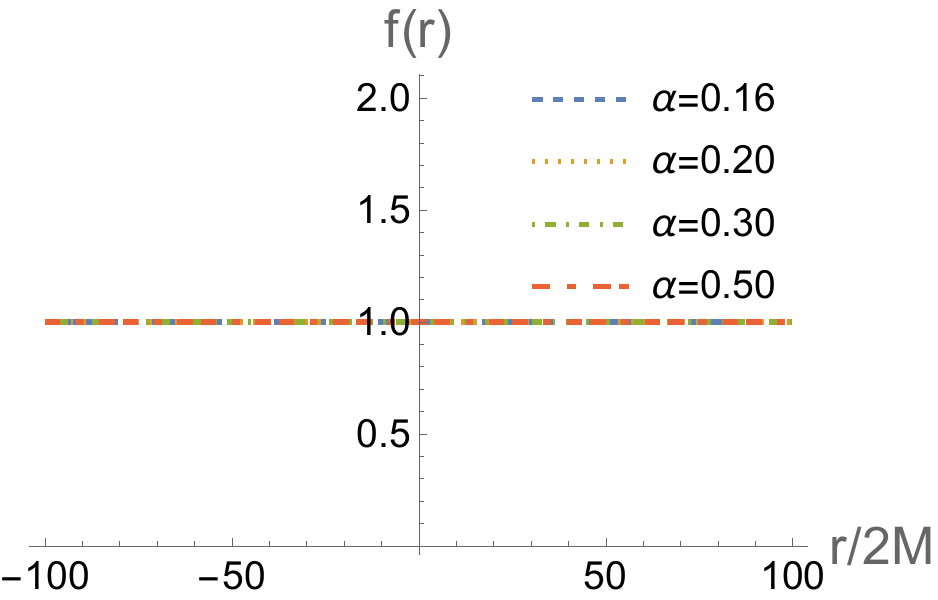}
}
{
\label{fig:b}
\includegraphics[width=0.3\columnwidth]{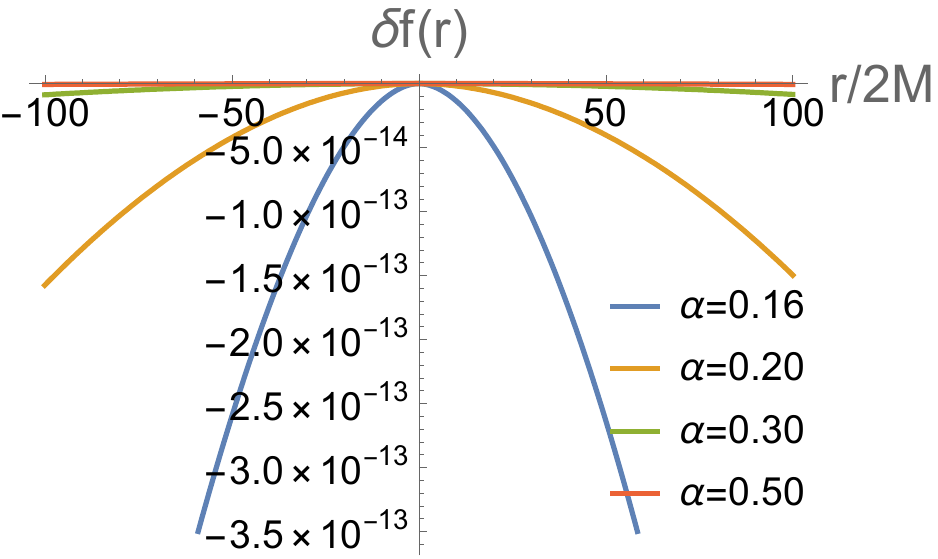}
}
\caption{The metric $f(r)$ of the Minkowski spacetime (left panel), pure dark matter spacetime and their differences $\delta f(r)$ (right panel) as a function of the spatial parameter $r$ with different shape parameters $\alpha$. The parameters we used are $ \rho_e=6.9 \times 10^6 M_\odot/kpc^3, r_e=91.2kpc$. We have converted these main calculation parameters to the black hole units before plotting.}
\label{f1}
\end{figure*}

In this section, we introduce an interesting and significant density profile of dark matter halo, that is the Einasto profile \cite{1965On}. Recently, J. Wang et al. indicated again that the structure of the dark matter halo can be well described by the Einasto profile using the parameters like the density of halo and characteristic radius \cite{Wang:2019ftp}
\begin{equation}
\rho_{\text{eina}} (r)=\rho_e \exp [ -2 \alpha ^{-1} ((r/r_e)^\alpha -1 ) ],
\label{e21}
\end{equation}
where, $r_e$ is the characteristic radius, $\rho_e$ is the density parameter and $\alpha$ is the shape parameter of dark matter halo. Since the suggested values for the shape parameter $\alpha$ in the density profile are different in Refs. \cite{Udrescu:2018hvl,Wang:2019ftp,Quintana:2022yky,Figueiredo:2023gas,Rahaman:2023tkm}, we reserve the parameter $\alpha$ as a free parameter and participate in the following calculations. Then, using this density profile in Eq.(\ref{e21}), the mass distribution of the dark matter halo can be calculated as follows
\begin{equation}
\begin{aligned}
M_{\text{eina}}(r)=4\pi \int_{0}^{r}\rho_{\text{eina}} (r')r'^2dr' =\frac{4\pi e^{2/\alpha } \rho_e \left ( 8^{-1/\alpha } r_e^3 \alpha ^{3/\alpha }\Gamma [\frac{3}{\alpha }] -r^3 \text{Ei}[\frac{-3 + \alpha}{\alpha },\frac{2}{\alpha }(\frac{r}{r_e})^\alpha] \right ) }{\alpha },
\label{e22}
\end{aligned}
\end{equation}
where, $\text{Ei}[n,z]$ denotes the exponential integral function $\text{Ei}_n(z)$ and $\Gamma[z]$ denotes the gamma function $\Gamma(z)$. In the Newtonian theory, when a test particle located at the equatorial plane of spherical symmetric spacetime, the tangential velocity of this particle can be determined by the mass distribution of dark matter halo \cite{Matos:2003nb,Jusufi:2019nrn}. So, this tangential velocity $V$ can be defined as
\begin{equation}
\begin{aligned}
V_{\text{eina}}(r)=\sqrt{\frac{M_\text{eina}}{r}}= \sqrt{4\pi e^{2/\alpha } \rho_e \left ( 8^{-1/\alpha } r_e^3 \alpha ^{3/\alpha }\Gamma [\frac{3}{\alpha }]/r -r^2 \text{Ei}[\frac{-3 + \alpha}{\alpha },\frac{2}{\alpha }(\frac{r}{r_e})^\alpha] \right )/\alpha  }.
\label{e23}
\end{aligned}
\end{equation}
On the other hand, if we consider a spherically symmetric spacetime full of the dark matter, the geometry of this spacetime can generally be given by
\begin{equation}
\begin{aligned}
ds^{2}=-f(r)dt^{2}+\frac{1}{g(r)}dr^{2}+H(r)(d\theta ^{2}+\sin^{2}\theta d\phi ^{2}),
\label{e24}
\end{aligned}
\end{equation}
where, $f(r)$ stands for the redshift function, $g(r)$ stands for the shape function. In the Ref. \cite{Matos:2003nb}, T. Matos et al. showed that this tangential velocity can be expressed as a function of metric coefficient $f(r)$, that is
\begin{equation}
V_\text{eina}^2(r)=\frac{r}{\sqrt{f(r)}}\frac{d\sqrt{f(r)}}{dr}=r\frac{dln\sqrt{f(r)}}{dr}
\label{e25}
\end{equation}
Here, we mainly consider the simplest case in Eq. (\ref{e24}), that is $f(r)=g(r)$. Then based on Eqs. (\ref{e23}) and (\ref{e25}), we can obtain an analytical solution of the metric coefficient $f(r)$ in a dark matter halo, that is
\begin{equation}
f_{\text{eina}}(r)=g_{\text{eina}}(r)=\exp (- \frac{4\pi 2^{1-3/\alpha } e^{2/\alpha } r^2 ((r/r_e)^{\alpha }/\alpha )^{-3/\alpha }\rho_e \Gamma [\frac{3}{\alpha },0,\frac{2(\frac{r}{r_e})^\alpha} {\alpha} ]}{\alpha}   ),
\label{e26}
\end{equation}
where, $\Gamma[a,z_0,z_1]$ denotes the generalized incomplete gamma function and it is given by the expression $\Gamma [a,z_0,z_1]=\int_{z_0}^{z_1}t^{a-1}e^{-t}dt$. At this point, combining Eqs. (\ref{e24}) and (\ref{e26}), the metric of pure dark matter space-time can be obtained. Here, we are first interested in the difference between the spacetime of pure dark matter and the Minkowski spacetime $(f(r)=1)$, which may help to qualitatively understand this spacetime of pure dark matter. Therefore, in this work, we consider the parameter of dark matter in the galaxy M87 and then compared it with the Minkowski spacetime. The halo density and characteristic radius in the galaxy M87 are $\rho_e=6.9 \times 10^6 M_\odot/kpc^3, r_e=91.2kpc$. Since the values of these parameters in the galaxy M87 are given in a non-natural unit system, converting them to a unified unit system will help the following discussion. Here, we convert the mass model of the galaxy M87 into the black hole unit system using the Schwarzschild black hole radius as the standard. The converting formulas are as follows: $r_\text{e}(\text{BHU})=r_\text{e} / (2 G M_{\text{BH}}/c^2) \times r_\text{BHU}$ and $\rho_\text{e}(\text{BHU})=\rho_\text{e} /(M_\text{BH}/(4/3 \pi (2 G M_{\text{BH}}/c^2)^3))\times \rho_{\text{BHU}}$\footnote{Here, please note that $M_\text{BH}$ refers to the actual mass of the black hole before conversion to the black hole unit, and $M$ is the mass of the black hole after the conversion. In the subsequent discussion, the mass we used all refer to the parameters converted to the black hole unit, that is $M$.}. Finally, with mass of the black hole $M(\text{BHU})=1/2$, we have the parameters of dark matter halo in black hole units, $\rho_\text{e}(\text{BHU})=\rho_\text{e} /(M_\text{BH}/(4/3 \pi (2 G M_{\text{BH}}/c^2)^3))\times \frac{M(\text{BHU})}{4/3\pi(2M(\text{BHU}))^3} \approx 1.41\times10^{-22}$, $r_\text{e}(\text{BHU})=r_\text{e} / (2 G M_{\text{BH}}/c^2) \times 2M(\text{BHU}) \approx 1.42 \times 10^8$. Then, we present the results in Figure \ref{f1}. From the left and middle panels, pure dark matter spacetime is very similar to Minkowski spacetime. In order to distinguish them accurately, in the right panel, we define their difference $\delta f(r)=f_{\text{eina}}(r)-1$, where the ``1'' represents the value of Minkowski spacetime. Firstly, from these results in the right panel, all the values of these differences are negative which means that the pure dark matter spacetime is lower than the Minkowski spacetime. Secondly, we find that for a specific shape parameter $\alpha$, the difference increases with the increasing of the spatial parameter $r$, and then decreases again. Besides, the most important finding is that we find these differences decrease with the increasing of the shape parameter $\alpha$, eventually leading to the phenomenon of zero difference. Zero difference indicates that dark matter spacetime and Minkowski spacetime are indistinguishable, which is obviously not allowed. Therefore, from this perspective, there is a theoretical upper limit to the shape parameter $\alpha$ of the Einasto profile. Here, based on the right panel, we give this upper limit of the shape parameter $\alpha$ and it is approximately $\alpha<0.50$. This result we obtained is basically consistent with the parameter used in Ref. \cite{Wang:2019ftp}. Of course, this is an ideal situation. In fact, the space of our universe is not empty or contains only a single object, and there are interactions between these objects. This upper limit makes more sense if one takes into account the interaction between dark matter and objects, such as the black hole or the compact object. Therefore, in the next section we will study the spacetime and properties of a black hole in a dark matter halo and re-estimate this upper limit. Finally, regarding the dark matter density and characteristic radius, it is not difficult to find that when dark matter is absent or the parameter $r$ tends to infinity, pure dark matter spacetime degenerates into Minkowski spacetime, that is
\begin{equation}
\lim_{\rho_\text{e}\rightarrow 0}f_{\text{eina}}(r)  = 1,\lim_{r\rightarrow \infty }f_{\text{eina}}(r) =1.
\label{e27}
\end{equation}

\section{One formal solution of black holes in a dark matter halo}\label{s3}
\begin{figure*}[t!]
\centering
{
\label{fig:b}
\includegraphics[width=0.3\columnwidth]{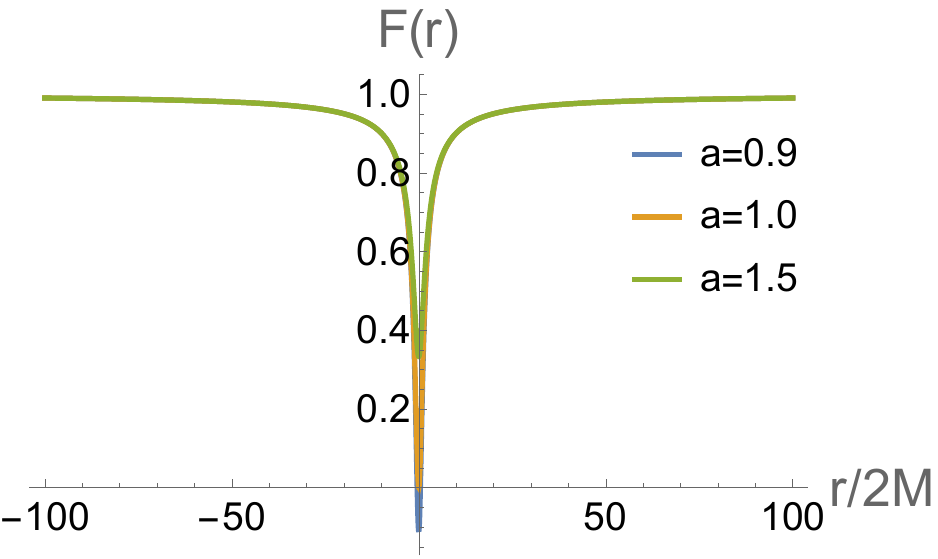}
}
{
\label{fig:b}
\includegraphics[width=0.3\columnwidth]{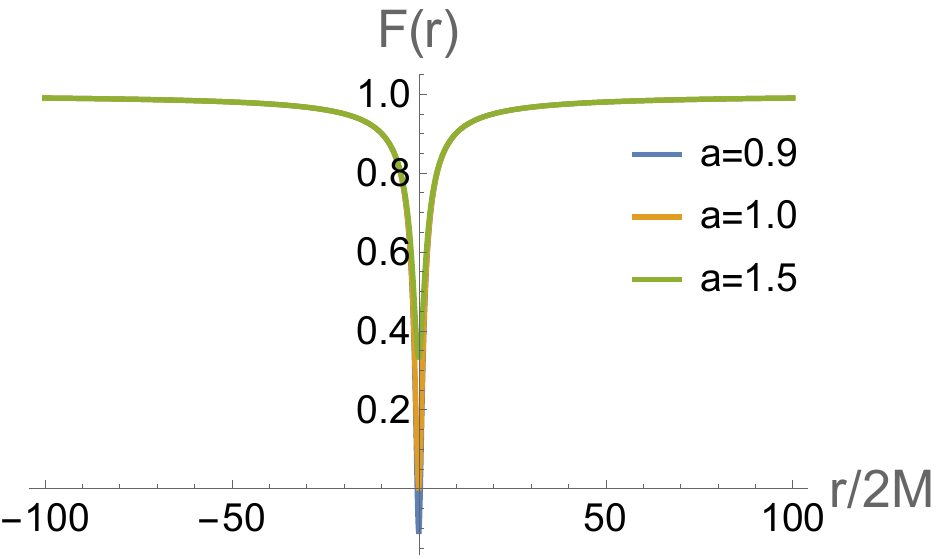}
}
{
\label{fig:b}
\includegraphics[width=0.3\columnwidth]{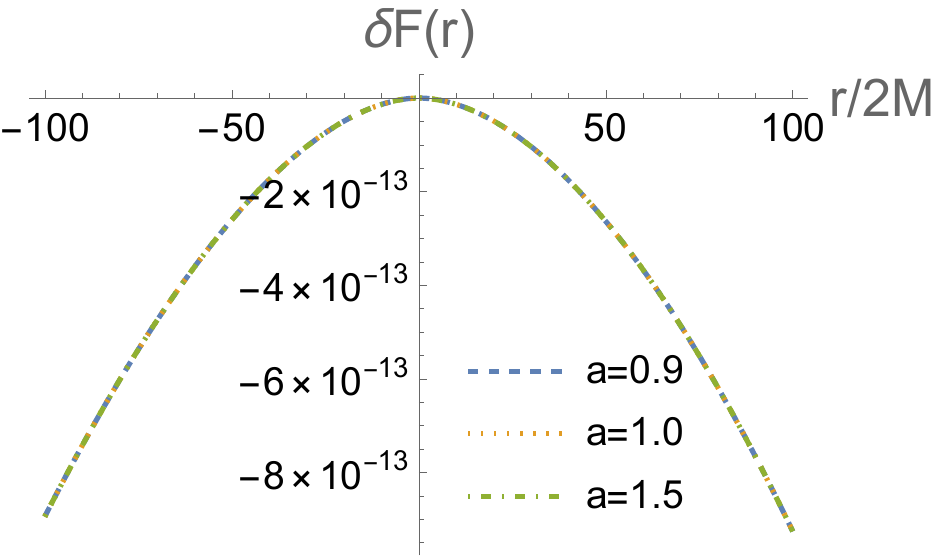}
}
\caption{The metric functions of the black bounce spacetime (left panel), dark matter spacetime and their differences (right panel) as a function of the spatial parameter $r$ with different geometric parameters $a$ in the M87. The parameters we used are $M=1/2, \alpha =0.16, \rho_e=6.9 \times 10^6 M_\odot/kpc^3, r_e=91.2kpc$. We have converted these main calculation parameters to the black hole units before plotting.}
\label{f2}
\end{figure*}

In this section, we extend our results (Eqs. (\ref{e24}) and (\ref{e26})) to the case of the black hole (BH) in a dark matter halo. With the methods recorded in Refs. \cite{Xu_2018,Xu:2021dkv}, these authors all show that, in Einstein field equations, the energy-momentum tensor of a black hole in a dark matter halo can usually be considered as a linear combination of the vacuum tensor and pure dark matter tensor. Therefore, we firstly need to solve the energy-momentum tensor of pure dark matter spacetime, and it satisfy the Einstein field equations
\begin{equation}
R_{\mu \nu }-\frac{1}{2}g_{\mu \nu }R=\kappa  T_{\mu \nu }(\text{DM-halo}),
\label{en9}
\end{equation}
where, $\kappa=8\pi$, $R_{\mu \nu }$, $R$ are Ricci tensor and Ricci scalar. $T_{\mu\nu}$ is the energy-momentum tensor of dark matter halo and it can be defined as ${T^\nu}_\mu=g^{\nu \alpha }T_{\mu \alpha }=diag[-\rho ,p_r, p, p]$. Taking the metric (\ref{e24}) into Eq. (\ref{en9}), we can obtain
\begin{equation}
\begin{aligned}
&\kappa  {T^{t}}_{t}(\text{DM-halo})=g(r)\left (\frac{H'}{2H}\frac{g'(r)}{g(r)}+\frac{H''}{H}-\frac{H'^2}{4H^2} \right ) - \frac{1}{H(r)}, \\
&\kappa  {T^{r}}_{r}(\text{DM-halo})=g(r)\left ( \frac{H'}{2H}\frac{f'(r)}{f(r)}+ \frac{H'^2}{4H^2} \right )-\frac{1}{H(r)},\\
&\kappa  {T^{\theta }}_{\theta }(\text{DM-halo})=\kappa  {T^{\phi }}_{\phi }(\text{DM-halo})\\
&=\frac{1}{2}g(r)\left ( \frac{f''(r)f(r)-f'(r)^2/2}{f(r)^2}+\frac{H'(r)}{2H(r)}(\frac{f'(r)}{f(r)}+\frac{g'(r)}{g(r)}) + \frac{f'(r)g'(r)}{2f(r)g(r)} +\frac{H''(r)}{H(r)}-\frac{H'(r)^2}{2H(r)^2}\right ).
\label{en10}
\end{aligned}
\end{equation}
where, functions $f(r)$, $g(r)$ can be obtained from Eqs. (\ref{e24}) and (\ref{e26}). If considering the black hole in a dark matter halo, the Einstein field equations can be rewritten as
\begin{equation}
R_{\mu \nu }-\frac{1}{2}g_{\mu \nu }R=\kappa  (T_{\mu \nu }(\text{BH})+T_{\mu \nu }(\text{DM-halo})).
\label{e31}
\end{equation}
where, $R_{\mu \nu }$ stands for Ricci tensor, $g_{\mu \nu }$ stands for the metric of a black hole in a dark matter halo and $R$ stands for Ricci scalar. Similarly, for a black hole in a dark matter halo, its geometric metric can be written as a linear combination of black hole and dark matter,
\begin{equation}
\begin{aligned}
ds^{2}=-(f(r)+F1(r))dt^{2}+\frac{1}{g(r)+G1(r)}dr^{2}+H(r)(d\theta ^{2}+\sin^{2}\theta d\phi ^{2}),
\label{e32}
\end{aligned}
\end{equation}
where, $f(r)$ and $g(r)$ stand for the metric coefficients of the pure dark matter halo in Eq. (\ref{e26}). Now, with the help of Eq. (\ref{en9}), taking Eq. (\ref{e32}) into Eq. (\ref{e31}) and using $T_{\mu \nu }(\text{BH})=0$ for the vacuum, we can always obtain the following forms from the component of the Einstein field equations
\begin{equation}
\begin{aligned}
&(g(r)+G1(r))\left ( \frac{H'}{2H}\frac{g'(r)+G1'(r)}{g(r)+G1(r)}+\frac{H''}{H}-\frac{H'^2}{4H^2} \right ) =g(r)\left (\frac{H'}{2H}\frac{g'(r)}{g(r)}+\frac{H''}{H}-\frac{H'^2}{4H^2} \right ),\\
&(g(r)+G1(r))\left ( \frac{H'}{2H}\frac{f'(r)+F1'(r)}{f(r)+F1(r)}+ \frac{H'^2}{4H^2} \right )=g(r)\left ( \frac{H'}{2H}\frac{f'(r)}{f(r)}+ \frac{H'^2}{4H^2} \right ),
\label{e33}
\end{aligned}
\end{equation}
where, $H$ denotes $H(r)$, $H'$ denotes $dH(r)/dr$ and $H''$ denotes $d^2H(r)/dr^2$. Here, Eq. (\ref{e33}) can be described by the combination of two different Einstein field equations components in Eqs. (\ref{en9}) and (\ref{e31}). In Eq. (\ref{e33}), Only the function $G1(r)$ in the first equation is unknown if $H(r)$ can be uniquely determined. The second equation is based on the functions $G1(r), F1(r), H(r)$ and it can be simplified as follows,
\begin{equation}
\begin{aligned}
\frac{f'(r)+F1'(r)}{f(r)+F1(r)} = \frac{g(r)}{g(r)+G1(r)}\left ( \frac{f'(r)}{f(r)}+ \frac{H'}{2H} \right ) - \frac{H'}{2H}.
\label{e34}
\end{aligned}
\end{equation}
The general solution for the first equation of Eq. (\ref{e33}) is $G1(r)=\frac{C\sqrt{H(r)}}{H'(r)^2}$, where, $C$ is an undetermined value, which can be obtained by the boundary condition. Different from the Schwarzschild black hole introduced in the Refs. \cite{Xu_2018,Jusufi:2019nrn,Xu:2021dkv}  as the boundary condition at infinity, here, we introduce a new special spacetime boundary condition for $G1(r)$ and $ H(r)$. In consideration of Eq.(\ref{e27}), these impacts of dark matter on spacetime at infinity can be equivalent to the black bounce spacetime. Therefore, we use the black bounce spacetime as the boundary condition at infinity. The metric of this black bounce spacetime (BBS) generally has the following form, that is $F_\text{BBS}(r)=1-2M/\sqrt{r^2+a^2}$ and $H_\text{BBS}(r)=r^2+a^2$ \cite{Simpson:2018tsi}. This black bounce spacetime has been extensively studied as it could imply the transformation of regular black holes into wormholes \cite{Simpson:2019cer,Lobo:2020ffi,Tsukamoto:2020bjm,Ou:2021efv,Yang:2021cvh,Guo:2021wid,Xu:2021lff,Ghosh:2022mka,AbhishekChowdhuri:2023ekr}. Based on this boundary condition, the new metric $G(r)=g(r)+G1(r)$ of the black hole in a dark matter halo will degenerate into black bounce spacetime at its boundary ($G(r)=1-2M /\sqrt{r^2+a^2}$). Therefore, a particular solution that satisfies the boundary condition is $g(r)=1$, $G1(r)=-2M/\sqrt{r^2+a^2}$. Finally, we can obtain the value $C$ and take it back to the general solution, a solution of $G1(r)$ can be obtained. Therefore, the analytical expression of function $G1(r)$ and $ F1(r)$ are given by
\begin{equation}
\begin{aligned}
&G1(r)= -\frac{2M}{\sqrt{r^2+a^2}},\\
&F1(r)=\exp\left ( \int \frac{g(r)}{g(r)+G1(r)}\left ( \frac{f'(r)}{f(r)}+ \frac{H'}{2H}  \right )dr- \frac{H'}{2H}dr \right )-f(r),
\label{e35}
\end{aligned}
\end{equation}
where, $M$ is the mass of the black hole and $a$ is a geometric parameter of the spacetime. Using the assumption of the previous section $f(r)=g(r)$ in Eq. (\ref{e33}), it is not difficult to find that $F1(r)=G1(r)= - 2M / \sqrt{r^2+a^2}$ only needing
\begin{equation}
\begin{aligned}
\frac{H''(r)}{H(r)}-\frac{H'(r)^2}{4H(r)^2}=\frac{H'(r)^2}{4H(r)^2}.
\label{e36}
\end{aligned}
\end{equation}
For the boundary condition of the black bounce spacetime, $H(r)=r^2+a^2$ is still satisfied with Eq. (\ref{e36}). Therefore, the new metric for the a black hole in a dark matter halo is given by
\begin{equation}
\begin{aligned}
ds^{2}=-F(r)dt^{2}+\frac{1}{G(r)}dr^{2}+H(r)(d\theta ^{2}+\sin^{2}\theta d\phi ^{2}),
\label{e37}
\end{aligned}
\end{equation}
where, $F(r)=f(r)+F1(r), G(r)=g(r)+G1(r), H(r)=r^2+a^2$, $f(r)$ and $g(r)$ represent the factor terms for considering pure dark matter halo in Eq. (\ref{e26}). Finally, the metric coefficients for a black hole in a dark matter halo are as follows
\begin{equation}
\begin{aligned}
F(r)&=G(r)=\exp (- \frac{4\pi 2^{1-3/\alpha } e^{2/\alpha } r^2 ((r/r_e)^{\alpha }/\alpha )^{-3/\alpha }\rho_e \Gamma [\frac{3}{\alpha },0,\frac{2(\frac{r}{r_e})^\alpha} {\alpha} ]}{\alpha}  ) - \frac{2M}{\sqrt{r^2+a^2}},
\label{e38}
\end{aligned}
\end{equation}
where, $a$ is geometric parameter of the spacetime and it is a non-negative number. Generally speaking, the different geometric parameters $a$ denote the different spacetimes of black hole in a dark matter halo: \romannumeral1) Schwarzschild-like black hole in a dark matter halo for the geometric parameter $a=0$; \romannumeral2) Regular black hole in a dark matter halo for the geometric parameter $0<a <2M$; \romannumeral3) The one-way wormhole in a dark matter halo for the geometric parameter $a=2M$; \romannumeral4) The traversable wormhole in the dark matter halo is for the geometric parameter $a>2M$. Where, the value $2M$ is an estimated threshold of the geometric parameter, that is $a_c = 2M$. In other words, when the geometric parameter $a>a_c$, the transformation from a regular black hole to a traversable wormhole can be achieved. Here, the event horizon $r_h$ of black holes can be obtained from Eq. (\ref{e38}), that is the positive real root of $F(r)=0$. Similarly, if the dark matter is absent, that is $\rho_\mathrm{e}=0$, we find that the Eq. (\ref{e38}) degenerates into the black bounce spacetime, that is $F(r)=G(r)=1-2M/\sqrt{r^2+a^2}$. If the case of $a=0$ is satisfied at this time, the spacetime will further degenerate into the  Schwarzschild black hole, that is $F(r)=G(r)=1-2M/r$. Besides, in Figure \ref{f2}, we show images of a black hole in a dark matter halo and black bounce spacetime as a function of spatial parameter $r$ with different geometric parameter $a$. From the left and middle panels, we find that the metric coefficient of black bounce spacetime (BBS) and dark matter (DM) spacetime are very similar and close. Bear similar to the analysis of the pure dark matter spacetime, by defining the difference $\delta F(r) = F(r)(\text{DM}) - F(r)(\text{BBS})$ in the right panel, our results show that, in the galactic center of M87, the metric coefficient $F(r)$ of dark matter spacetime is basically lower than the black bounce spacetime with the increasing of spatial $r$. This result is similar to our previous results considering pure dark matter spacetime in the previous section.
\begin{figure*}[t!]
\centering
{
\label{fig:b}
\includegraphics[width=0.3\columnwidth]{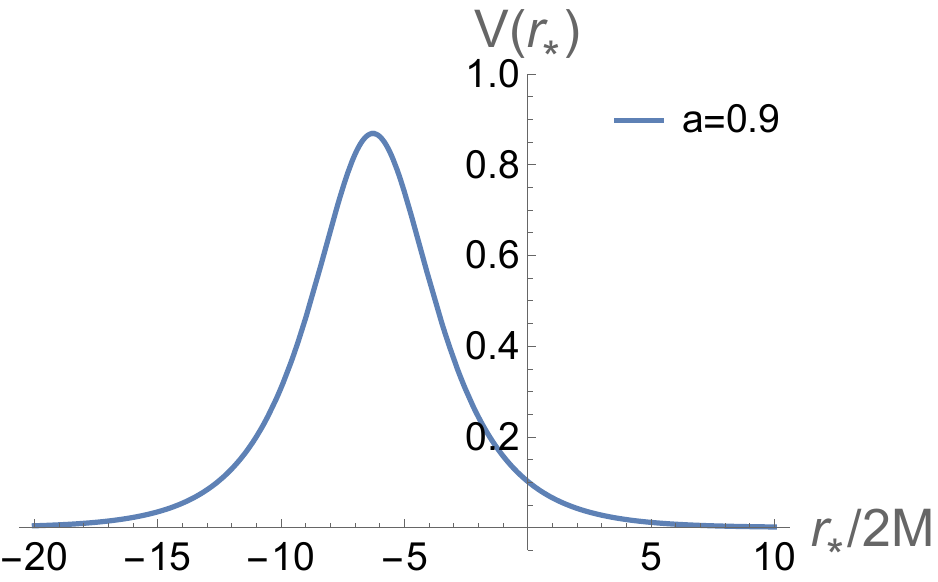}
}
{
\label{fig:b}
\includegraphics[width=0.3\columnwidth]{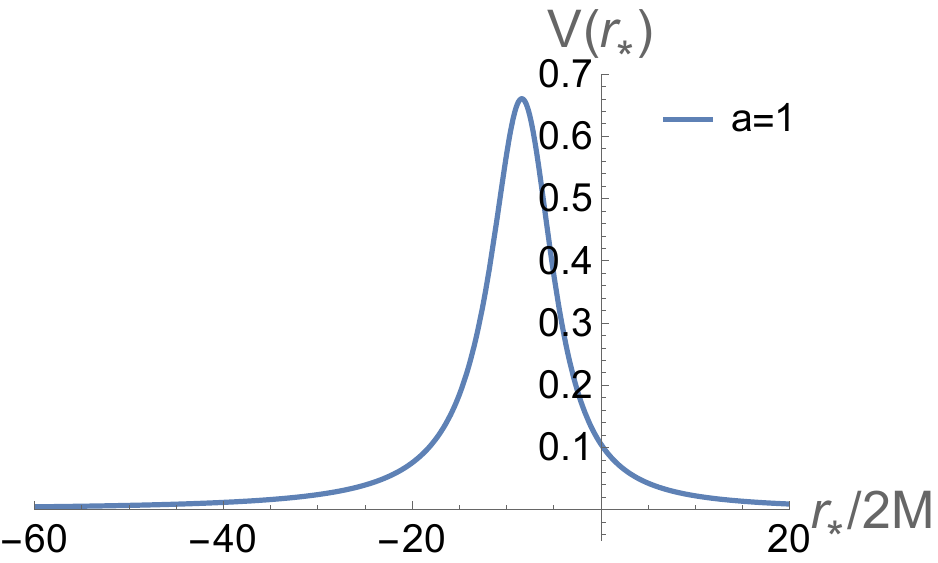}
}
{
\label{fig:b}
\includegraphics[width=0.3\columnwidth]{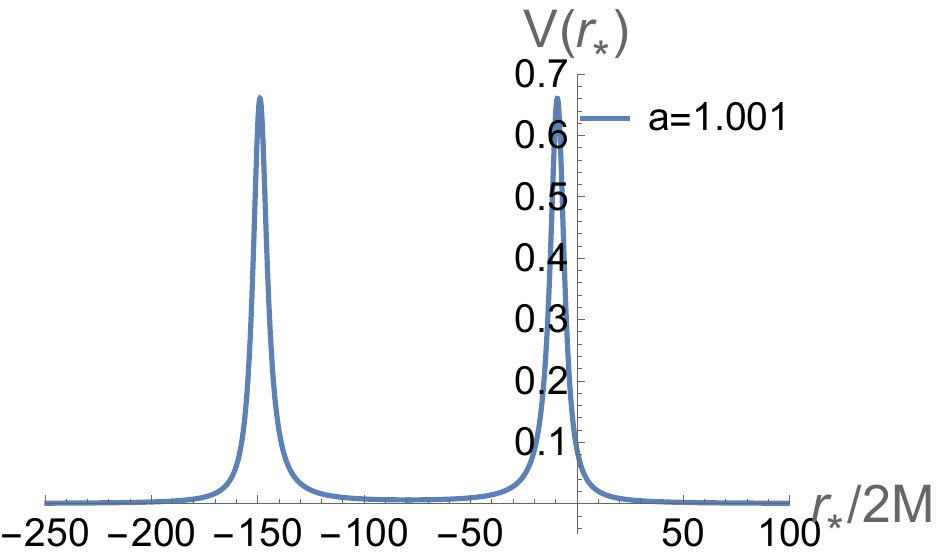}
}
{
\label{fig:b}
\includegraphics[width=0.3\columnwidth]{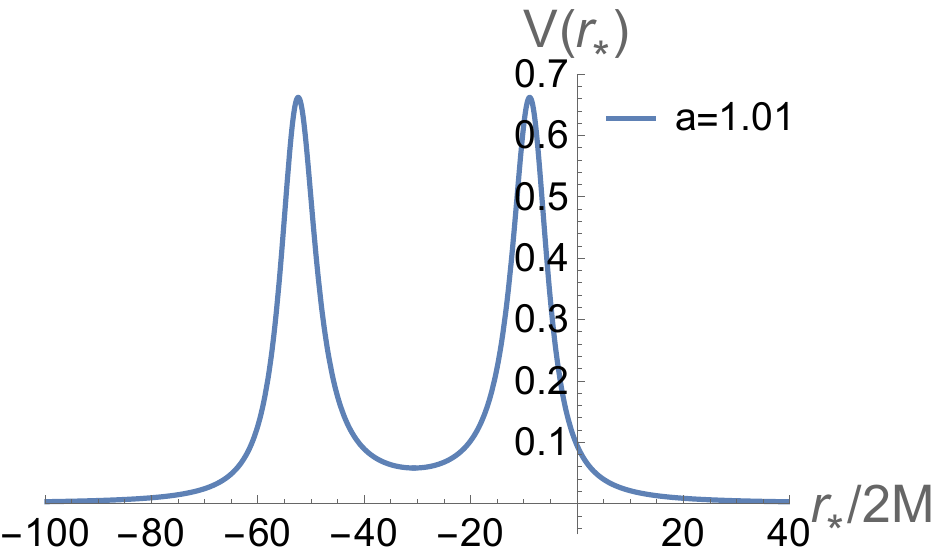}
}
{
\label{fig:b}
\includegraphics[width=0.3\columnwidth]{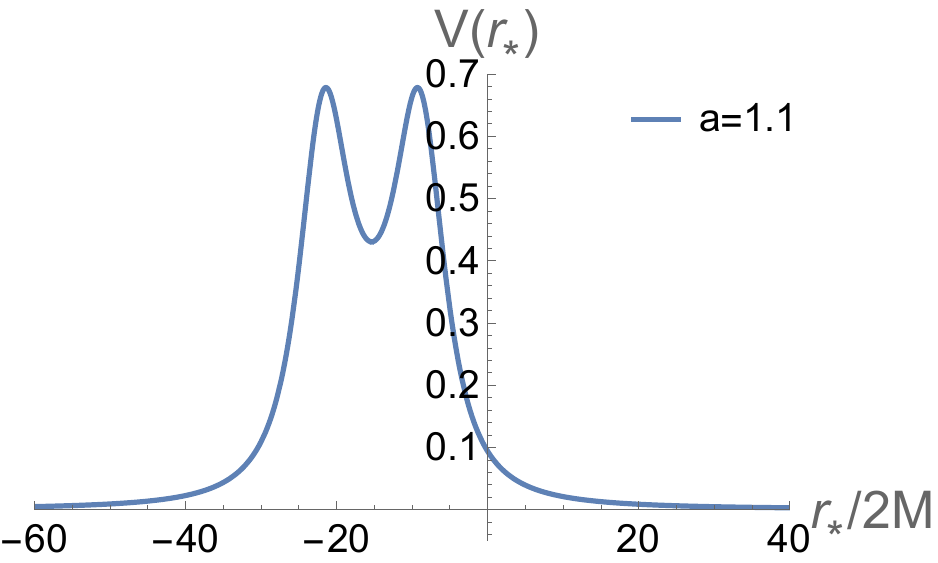}
}
{
\label{fig:b}
\includegraphics[width=0.3\columnwidth]{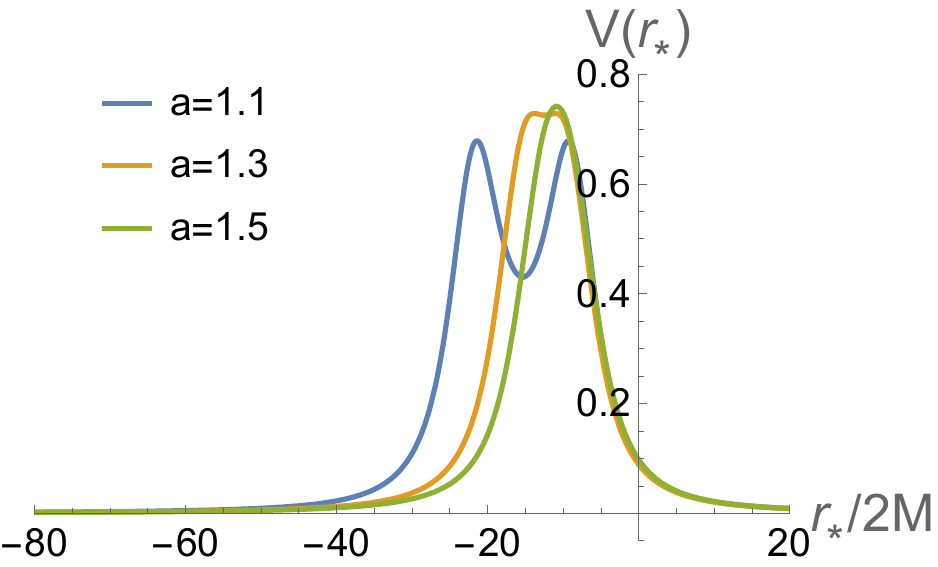}
}
\caption{The effective potentials of black holes as a function of the tortoise coordinates $r_*$ with different geometric parameters $a$ in a dark matter halo of M87. The parameters we used are $M=1/2, \alpha =0.16, \rho_e=6.9 \times 10^6 M_\odot/kpc^3, r_e=91.2kpc$. We have converted these main calculation parameters to the black hole units before plotting.}
\label{f3}
\end{figure*}
\begin{figure*}[t!]
\centering
{
\label{fig:b}
\includegraphics[width=0.45\columnwidth]{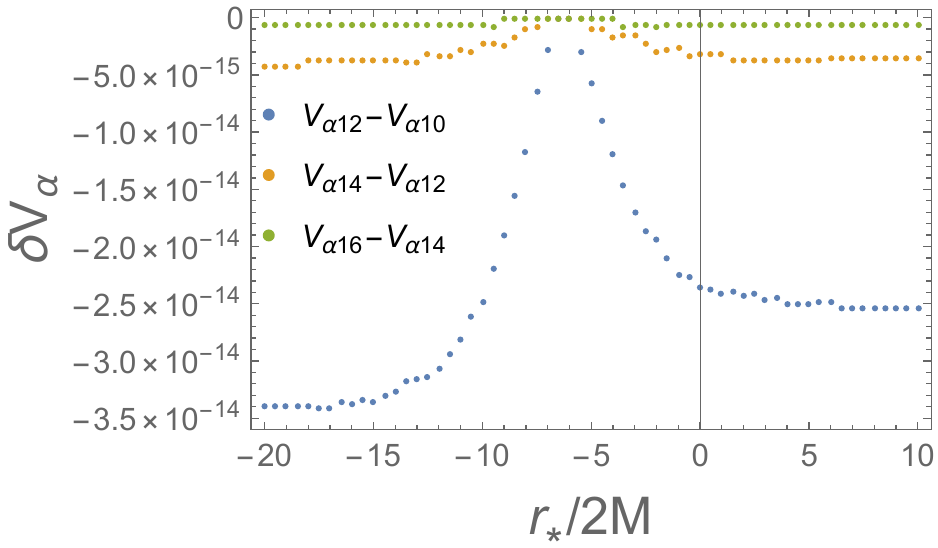}
}
{
\label{fig:b}
\includegraphics[width=0.45\columnwidth]{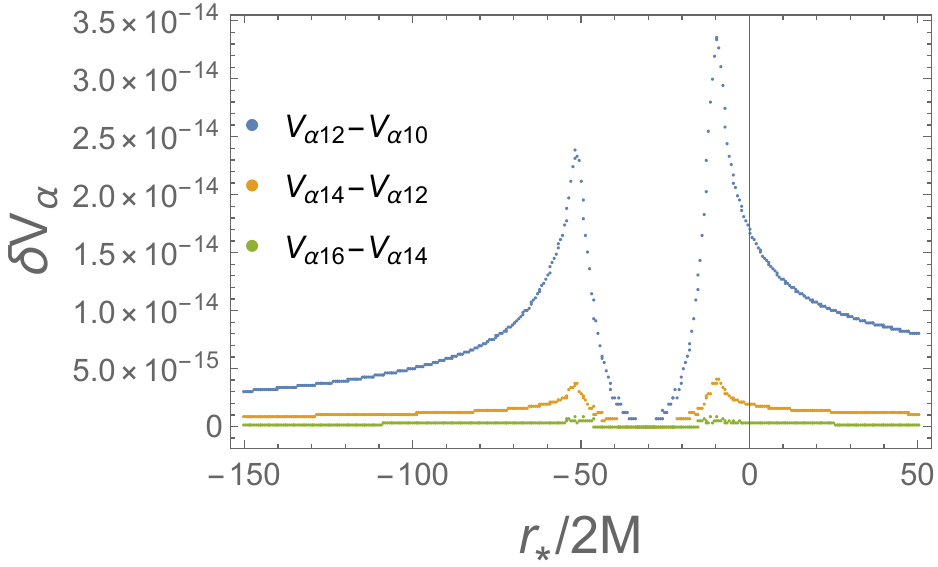}
}
\caption{The difference of the effective potential as a function of the tortoise coordinates $r_*$ in the geometric parameter $a=0.9$ (left panel) and $a=1.01$ (right panel) with different shape parameter $\alpha$ under the dark matter halo of M87. The parameters we used are $M=1/2, \rho_e=6.9 \times 10^6 M_\odot/kpc^3, r_e=91.2kpc$ and the step size of each point is 0.5. We have converted these main calculation parameters to the black hole units before plotting.}
\label{f4}
\end{figure*}
\section{Axial gravitational perturbation of black holes in a dark matter halo}\label{s4}
In this section, we mainly consider axial gravitational perturbation of the black holes in a dark matter halo. In the previous section, the spacetime metric of the black holes in a dark matter halo is given by Eq. (\ref{e37}) and the non-zero component of this metric can be written as
\begin{equation}
\bar{g}_{\mu\nu}=diag(-F(r),1/G(r), H(r), H(r)\sin^2 \theta).
\label{e41}
\end{equation}
where, $F(r), G(r)$ can be obtained in Eq. (\ref{e38}) and $H(r)=r^2+a^2$.  Gravitational perturbation, also known as metric perturbation, usually decomposes the metric into a background metric and a small perturbation. The background metric describes a stable spacetime, while the perturbation metric describes gravitational waves or other perturbation. Regge and Wheeler were the first to use gravitational perturbation to calculate the spacetime properties of the Schwarzschild black hole \cite{Regge:1957td}. These authors then used this method of gravitational perturbation to study the space-time properties of other black holes \cite{Ulhoa:2013fca,Cruz:2015bcj,Chowdhury:2020rfj}. Generally speaking, this gravitational perturbation can be divided into two types, namely axial perturbation and polar perturbation. Here, for simplicity, we mainly study the axial perturbation of the black hole in a dark matter halo. Strictly following the method in the above references, we can write the perturbed metric as
\begin{eqnarray}
g_{\mu \nu }=\bar{g}_{\mu \nu }+h_{\mu \nu },
\label{e42}
\end{eqnarray}
where, $\bar{g}_{\mu \nu }$ stands for the background metric of a black hole in a dark matter halo and $h_{\mu \nu }$ stands for the linear perturbation ($h_{\mu \nu } \ll \bar{g}_{\mu \nu }$) which reads
\begin{equation}
h_{\mu\nu}=\sum_{l=0}^{\infty }\sum_{m=-l}^{m=l}[(h_{\mu\nu}^{lm})^{\text{axial}}+(h_{\mu\nu}^{lm})^{\text{polar}}].
\label{e43}
\end{equation}
Similarly, based on this perturbed metric, the perturbed Christoffel symbols also can be rewritten as follows
\begin{eqnarray}
\Gamma _{\mu \nu }^{\lambda  }=\bar{\Gamma }_{\mu \nu }^{\lambda  }+\delta \Gamma _{\mu \nu }^{\lambda  },
\label{e44}
\end{eqnarray}
where, $\bar{\Gamma }_{\mu \nu }^{\alpha }$ stands for Christoffel symbols of the background metric and the $\delta \Gamma _{\mu \nu }^{\alpha }$ can be written as follows
\begin{eqnarray}
\delta \Gamma _{\mu \nu }^{\lambda  }=\frac{1}{2}\bar{g}^{\lambda  \beta }(h_{\mu \beta ;\nu} +h_{\nu \beta ;\mu } - h_{\mu \nu ;\beta }),
\label{e45}
\end{eqnarray}
where, the symbol of $;\nu$ stands for the covariant derivative to the background metric $\bar{g}_{\mu \nu }$. Then, the perturbed Ricci tensor can be written as follows
\begin{eqnarray}
R_{\mu \nu }=\bar{R}_{\mu \nu }+\delta R_{\mu \nu },
\label{e46}
\end{eqnarray}
where, $\bar{R}_{\mu \nu }$ stands for the Ricci tensor of the background metric and
\begin{eqnarray}
\delta R_{\mu \nu }=\delta \Gamma _{\mu \lambda  ;\nu  }^{\lambda}-\delta \Gamma _{\mu \nu;\lambda  }^{\lambda  },
\label{e47}
\end{eqnarray}
Due to the perturbed term $\delta R_{\mu \nu }$ of the Ricci tensor $R_{\mu \nu }$ has no contribution \cite{Kobayashi:2012kh,Chowdhury:2020rfj}. Therefore, the master equation of the axial perturbation can be given by
\begin{eqnarray}
\delta R_{\mu \nu }=0.
\label{e48}
\end{eqnarray}
In this work, we mainly consider the case of axial gravitational perturbation and the perturbed term $h_{\mu \nu}^{\text{axial}}$ can be written as follows \cite{Regge:1957td}
\begin{eqnarray}
h_{\mu \nu }^{\text{axial}}=\begin{pmatrix}
0 &  0& 0 & h_{0}(t,r)\\
 0& 0 &  0& h_{1}(t,r)\\
0 & 0 & 0 &0 \\
h_{0}(t,r) &h_{1}(t,r)  &0  & 0
\end{pmatrix}{\rm sin}\theta \partial \theta P_{l}({\rm cos}\theta )
\label{e49}
\end{eqnarray}
where $P_{l}({\rm cos}\theta )$ are the Legendre polynomials of the order $l$. Taking Eqs. (\ref{e41}) and (\ref{e49}) into Eq. (\ref{e48}), the perturbed term components $\delta R_{24}$ and  $\delta R_{34}$ can be written as
\begin{equation}
\begin{aligned}
&\frac{\partial^2}{\partial t ^2}h_1 -\frac{\partial}{\partial r}\frac{\partial}{\partial t}h_0 +\frac{H'}{H} \frac{\partial}{\partial t}h_0 - \frac{G F' H' - F [2 l (l+1) - G' H' - 2 G H'']}{2 H}h_1=0
\label{e410}
\end{aligned}
\end{equation}
and
\begin{equation}
\begin{aligned}
&-\frac{1}{F}\frac{\partial }{\partial t} h_0 +  G \frac{\partial}{\partial r}h_1 + \frac{F G' + G F'}{2 F}h_1=0,
\label{e411}
\end{aligned}
\end{equation}
where, the functions $h_0, h_1$ denote the $h_0(t, r), h_1(t,r)$ and the metric functions $F, G, H$ denote $F(r), G(r), H(r)$.
The prime symbols $'$ and $''$ on the metric functions mean the first and second derivative with respects to $r$. Then, using Eq. (\ref{e411}) to eliminate $h_0$ in Eq. (\ref{e410}) with $\psi (t,r)=\sqrt{F(r)G(r)/H(r)}h_1(t,r)$, we can obtain the following form
\begin{equation}
\begin{aligned}
\frac{\partial^2 }{\partial t^2} \psi(t,r) - FG\frac{\partial^2 }{\partial r^2} \psi(t,r) -\frac{GF'+FG'}{2} \psi(t,r) + V_\text{eff}(r) \psi(t,r)=0.
\label{e412}
\end{aligned}
\end{equation}
Finally, we introduce tortoise coordinates, which have the advantage of expanding the horizon to infinity. The tortoise coordinates can be defined by the equation below,
\begin{equation}
\begin{aligned}
dr_*=\frac{1}{\sqrt{F(r)G(r)}}dr.
\label{e413}
\end{aligned}
\end{equation}
After then we obtain the following form
\begin{equation}
\begin{aligned}
&\frac{\partial^2 }{\partial t^2} \psi(t,r) - \frac{\partial^2 }{\partial r_*^2} \psi(t,r) + V_\text{eff} (r) \psi(t,r)=0,
\label{e414}
\end{aligned}
\end{equation}
where,
\begin{equation}
\begin{aligned}
V_\text{eff} (r)=\frac{3FG(H')^2}{4H^2} &- \frac{GF'H' +FG'H'+2FGH''}{4H} \\
&- \frac{G F' H'-F [2 l (l+1)-G' H' -2G H'']}{2 H},
\label{e415}
\end{aligned}
\end{equation}
Similarly, for simplicity, we continue to use the assumption of the previous section in all the discussions that follow, that is $F(r)=G(r)$, $H(r)=r^2+a^2$, then $F(r)$ and $  G(r)$ can be obtained in Eq. (\ref{e38}). In particular, if dark matter is absent ($\rho_\text{e}=0$) and the geometric parameter tends to zero ($a=0$), the effective potential of Eq. (\ref{e415}) degenerates into the form of the Schwarzschild black hole, that is
\begin{equation}
\begin{aligned}
V_{\text{Sch}}(r)=(1-\frac{2M}{r})(\frac{l(l+1)}{r^2}-\frac{6M}{r^3}).
\label{e416}
\end{aligned}
\end{equation}
Here, $M$ is the mass of the Schwarzschild black hole. At this point, we have obtained the effective potential of a black hole in a dark matter halo under the axial gravitational perturbation. Similarly, we are interested in the impacts of dark matter parameters on the effective potential and their differences from the black bounce spacetime, which will help further our understanding of a black hole in a dark matter halo. In Figure \ref{f3}, we mainly show the effective potentials of a black hole both in the dark matter spacetime and the black bounce spacetime as a function of the tortoise coordinate $r_*$ under different geometric parameters $a$. Our results show that the maximum of the effective potential basically increases with increasing geometric parameter $a$. when the geometric parameter $a>2M$, the effective potential has an obvious double barrier feature. Then with the geometric parameter $a$ continuous increasing, the potential well between the double barrier rises and narrows, eventually returning to a single potential barrier. Besides, we also study the impacts of the shape parameter $\alpha$ in a dark matter halo on the effective potential. Therefore, in Figure \ref{f4}, we also show the effective potentials of a black hole in a dark matter halo as a function of the tortoise coordinate $r_*$ under different shape parameter $\alpha$. Considering that their differences may be very small, we define the difference of the effective potential between the different shape parameters $\alpha$, that is $\delta V_{\alpha(A-B)} (r_*)= V_{\alpha A} (r_*) - V_{ \alpha B} (r_*) $, where, $A, B$ are the different values of the shape parameter $\alpha$, respectively. Finally, our results show that under the same parameter, the effective potential decreases with the increasing of the shape parameter $\alpha$.

\section{Quasinormal modes and gravitational wave echoes of black holes in a dark matter halo}\label{s5}
\subsection{The time-domain method and the WKB method}
In order to obtain the time evolution and quasinormal modes (QNM) frequencies of the black holes in a dark matter halo under the gravitational perturbation, the methods we used are the time domain method and the WKB method. Firstly, strictly following the time domain method recorded in Refs. \cite{PhysRevD.49.883,Chowdhury:2020rfj}, using light-cone coordinates, that is $u=t-r_*$ and $v=t+r_*$, the master equation of the gravitaional perturbation in Eq. (\ref{e414}) can be written as follows
\begin{equation}
4\frac{\partial ^{2}\psi (u ,v )}{\partial u \partial v } + V_\text{eff}(u ,v)\psi (u ,v )=0,
\label{e51}
\end{equation}
where, $r_*$ in the ligh-cone coordinates are the tortoise coordinates. Eq. (\ref{e51}) is directly related to the effective potential of a black hole in a dark matter halo. For Eq. (\ref{e51}), a suitable integration grid recorded in Refs. \cite {PhysRevD.63.084014,PhysRevD.64.044024,PhysRevD.72.044027} can be discretized as
\begin{equation}
\Psi(N)=\Psi(W)+\Psi(E)-\Psi(S)-h ^2\frac{V(W)\psi(W)+V(E)\psi(E)}{8}+O(h ^4),
\label{e52}
\end{equation}
where, $h$ is the scale of each grid cell. The letters of the integration grid are $ N=(u+h, v+h)$, $W=(u+h, v)$, $E=(u, v +h)$ and $S = (u, v)$, respectively. The initial condition we used is the Gaussian wave packet \cite{Yang:2022ryf} for two null surface, that is $u = u_0$ and $v = v_0$,
\begin{equation}
\begin{aligned}
\psi (u=u _{0} ,v )&=A {\rm exp}\left [ -\frac{(v -v _{0})^{2}}{\sigma ^{2}} \right ],\\
\psi (u ,v=v_{0} )&=0,
\label{e53}
\end{aligned}
\end{equation}
where, $A=1$, $v_0=10$ and $\sigma=3$. In this way, we can obtain the time evolution of a black hole under the gravitational perturbation. Furthermore, based on the time evolution of the black hole under the gravitational perturbation, we can use the Prony method, fitting a signal of the quasinormal mode by superposition of damped exponents, to extract the frequencies of the quasinormal mode \cite{RevModPhys.83.793,PhysRevD.75.124017,Chowdhury:2020rfj},
\begin{eqnarray}
\psi(t)\simeq \sum ^p_{i=1}C_ie^{-i\omega_it}.
\label{e54}
\end{eqnarray}
Note that all the quasinormal mode frequencies we calculated in this work refer to the fundamental quasinormal mode frequencies, that is the case of $n=0$. This is because the signals of QNMs can be well approximated by fundamental mode, and the contribution of higher overtone can be neglectable \cite{Churilova:2019cyt}.

On the other hand, in this work, the effective potential of the axial gravitational perturbation can be divided into two types according to the geometric parameter $a$, that is, a single barrier for the parameter $a\leq 2M$ and a double barrier for the parameter $a>2M$. For a single barrier, a simple and effective method to calculate the QNM of the black hole is the WKB method. Therefore, we can also use the WKB method except the Prony method. This method was first proposed by Schutz and Will \cite{Schutz:1985km}, and then promoted by Iyer, Will, and Konoplya \cite{PhysRevD.35.3621,PhysRevD.35.3632,PhysRevD.68.024018,PhysRevD.68.124017,RevModPhys.83.793}. In order to obtain the QNM frequencies, we use the sixth-order WKB formula in Ref. \cite{PhysRevD.68.124017,PhysRevD.68.024018}:
\begin{eqnarray}
\frac{i (\omega ^{2}-V_{0})}{\sqrt{-2V_{0}^{''}}}-\sum ^{6}_{i=2}\Lambda _{i}=n+\frac{1}{2}, \quad(n=0,1,2, \cdots)
\label{e55}
\end{eqnarray}
where, $V_{0}$ is the maximum of the effective potential, the prime is the second derivative of $V$ at its maximum, $\Lambda _{i}$ is the $i$th order revision terms which can be obtained in Ref. \cite{PhysRevD.68.024018} and $n$ is the overtone number. In Eq. (\ref{e55}), the frequencies of quasinormal mode are related to the effective potential directly and they depend on the following parameters: $M$, $a$, $r_\text{e}$, $\rho_\text{e}$ and $\alpha$ . Therefore, we can study the impacts of quasinormal modes on the black hole or dark matter parameters. Similarly, in this work, we mainly calculate the frequencies of fundamental quasinormal mode, that is the case of $n=0$. In the next subsection, we will study and analyze these results one by one.
\begin{figure*}[t!]
\centering
{
\label{fig:b}
\includegraphics[width=0.31\columnwidth]{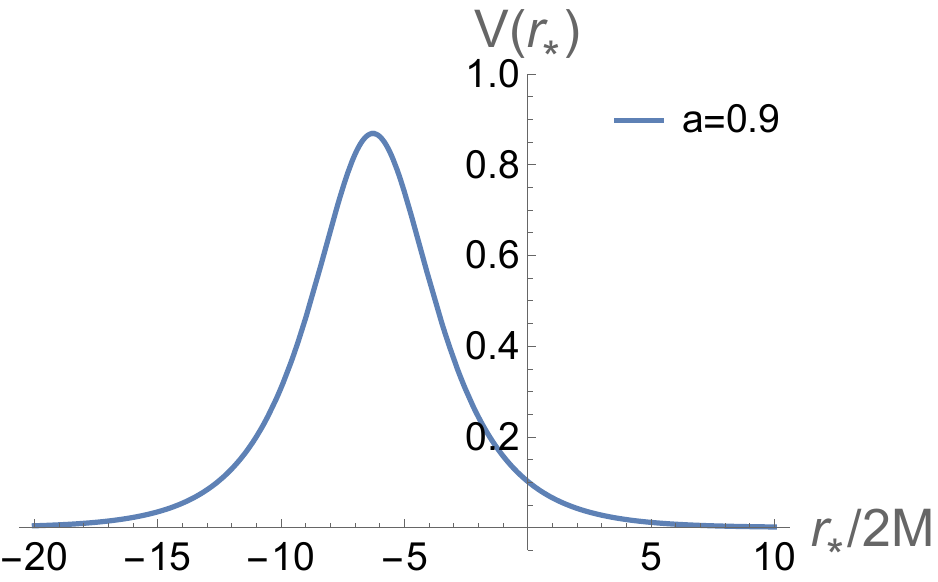}
}
{
\label{fig:b}
\includegraphics[width=0.31\columnwidth]{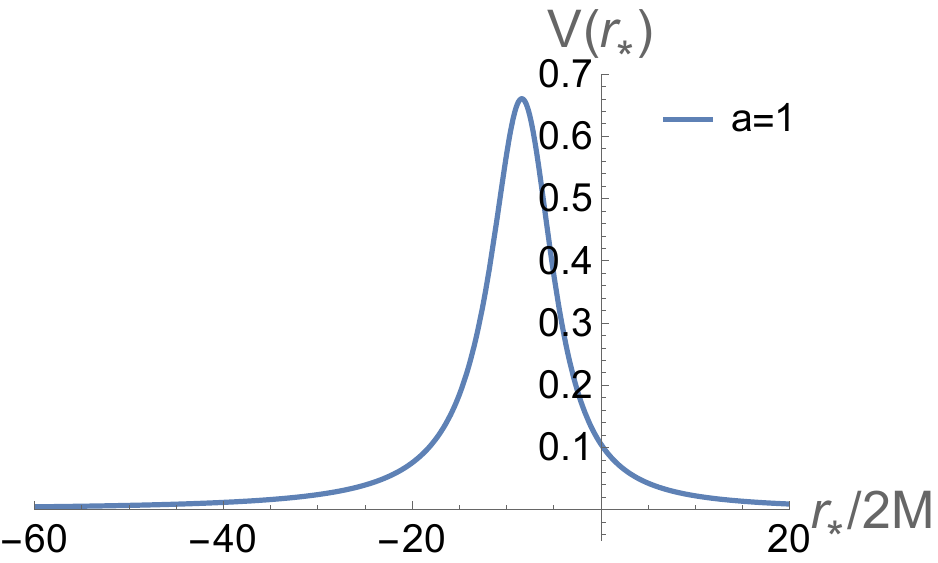}
}
{
\label{fig:b}
\includegraphics[width=0.31\columnwidth]{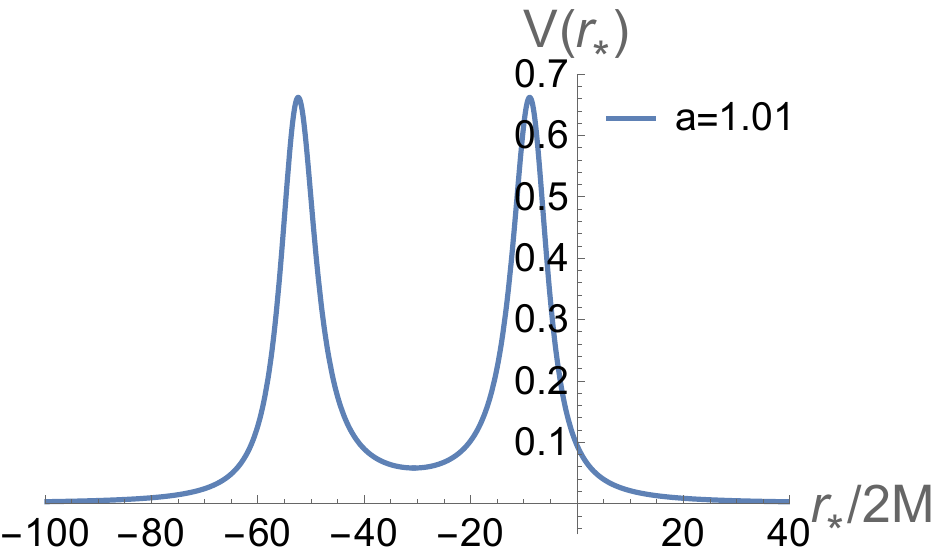}
}
{
\label{fig:b}
\includegraphics[width=0.31\columnwidth]{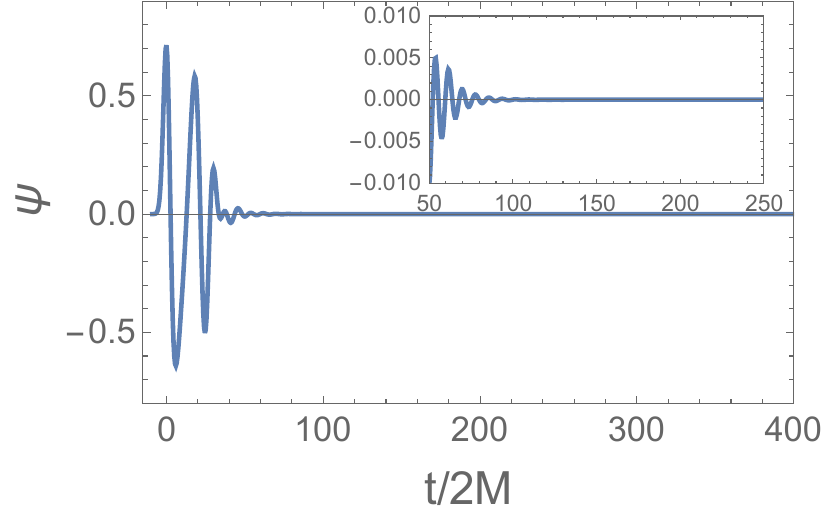}
}
{
\label{fig:b}
\includegraphics[width=0.31\columnwidth]{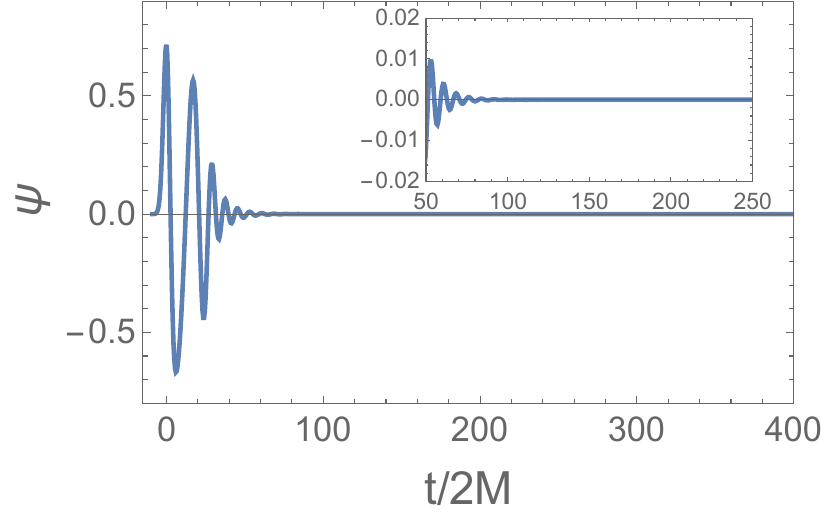}
}
{
\label{fig:b}
\includegraphics[width=0.31\columnwidth]{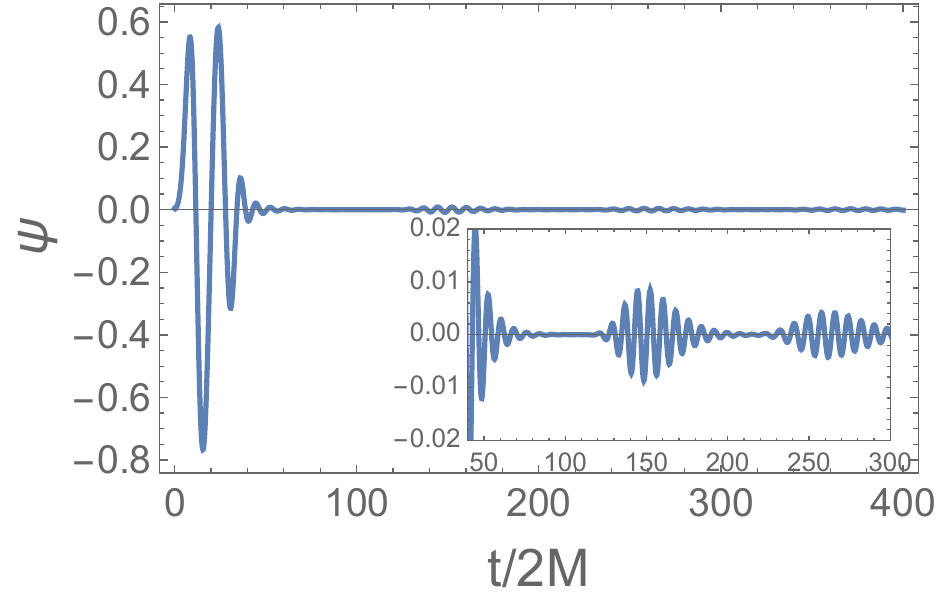}
}
\caption{The effective potentials of black holes as a function of the tortoise coordinates $r_*$ with different geometric parameter $a$ (top panels) and its time evolutions (bottom panels) under the dark matter halo of M87. The gravitational wave echoes emit when the geometric parameter is approximately $a>2M$. The parameters we used are $M=1/2, \alpha=0.16, \rho_e=6.9 \times 10^6 M_\odot/kpc^3, r_e=91.2kpc$. We have converted these main calculation parameters to the black hole units before plotting.}
\label{f5}
\end{figure*}
\begin{figure*}[t!]
\centering
{
\label{fig:b}
\includegraphics[width=0.31\columnwidth]{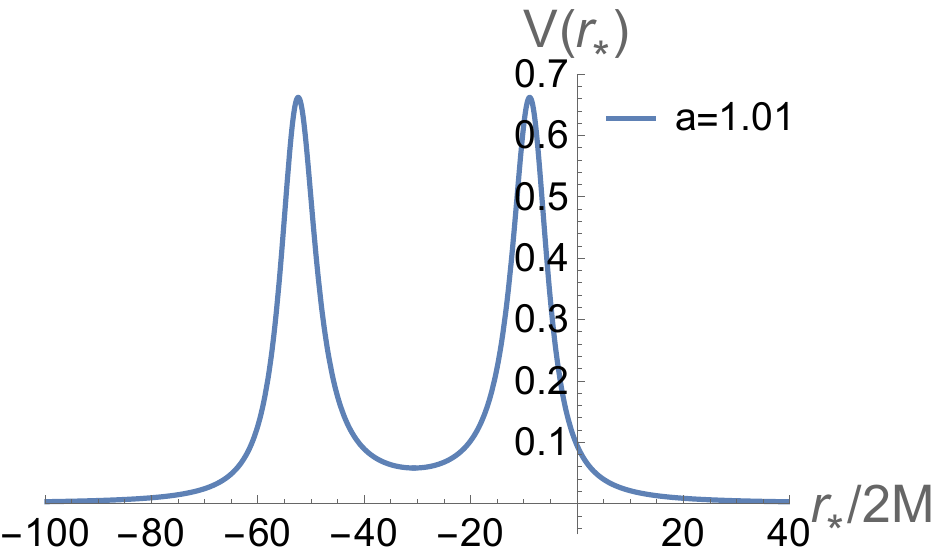}
}
{
\label{fig:b}
\includegraphics[width=0.31\columnwidth]{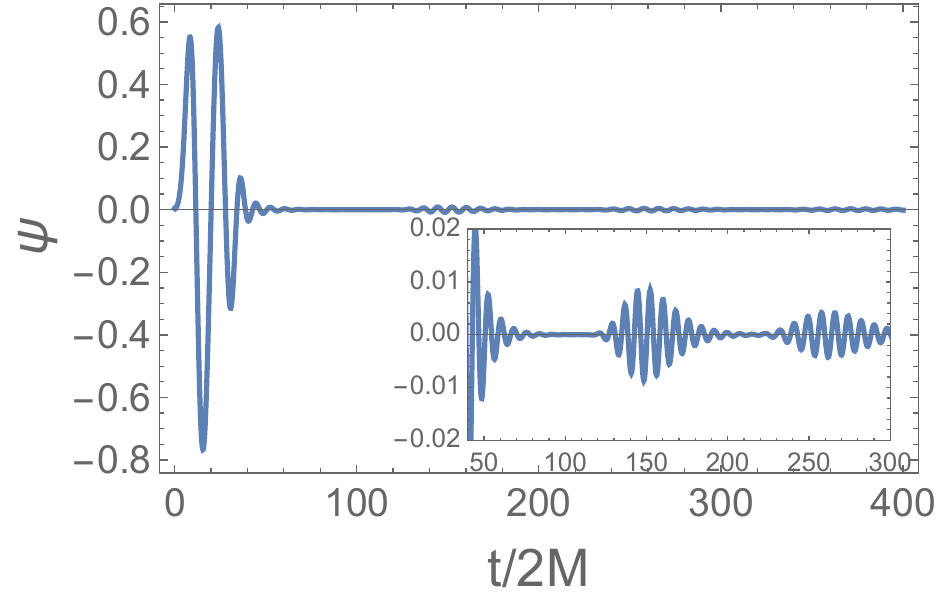}
}
{
\label{fig:b}
\includegraphics[width=0.31\columnwidth]{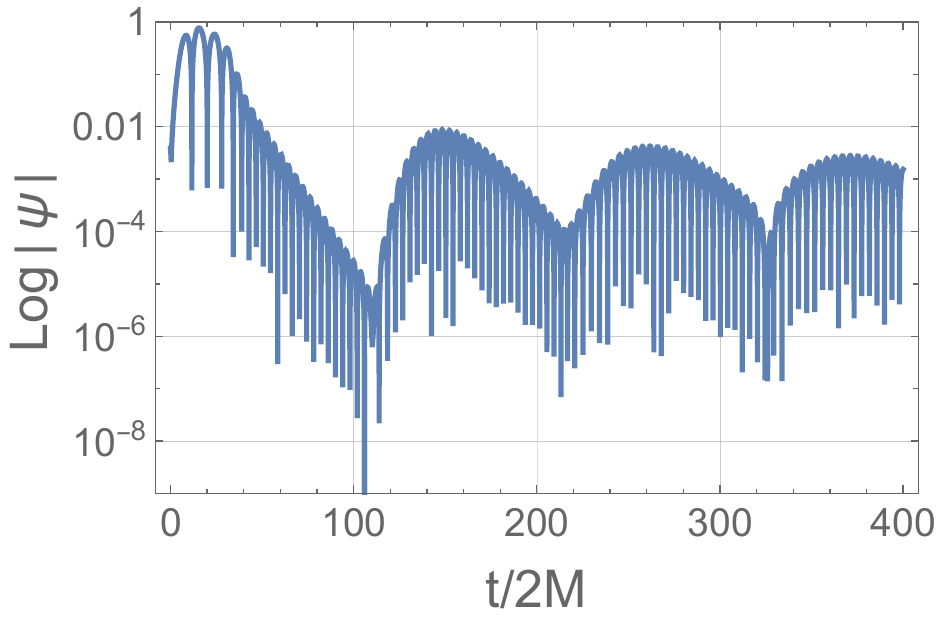}
}
{
\label{fig:b}
\includegraphics[width=0.31\columnwidth]{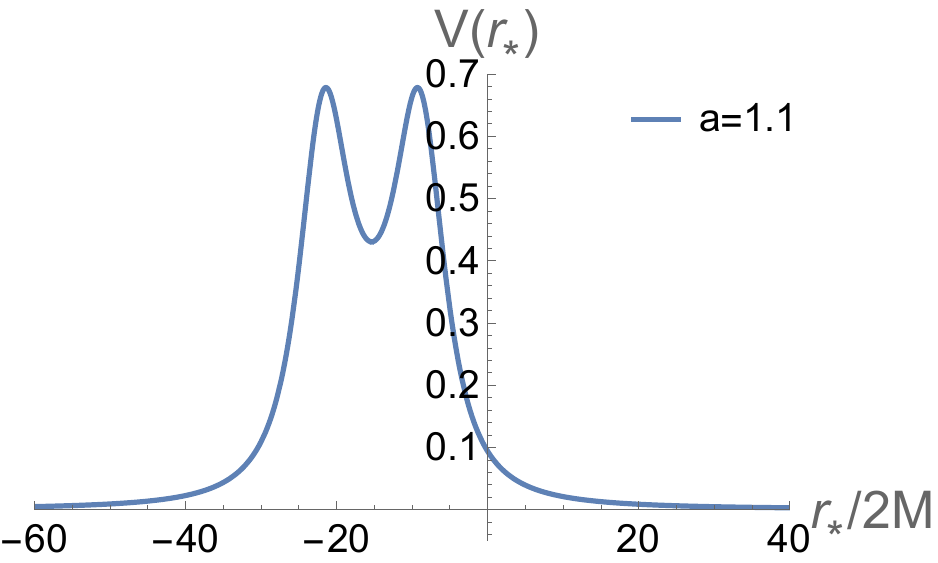}
}
{
\label{fig:b}
\includegraphics[width=0.31\columnwidth]{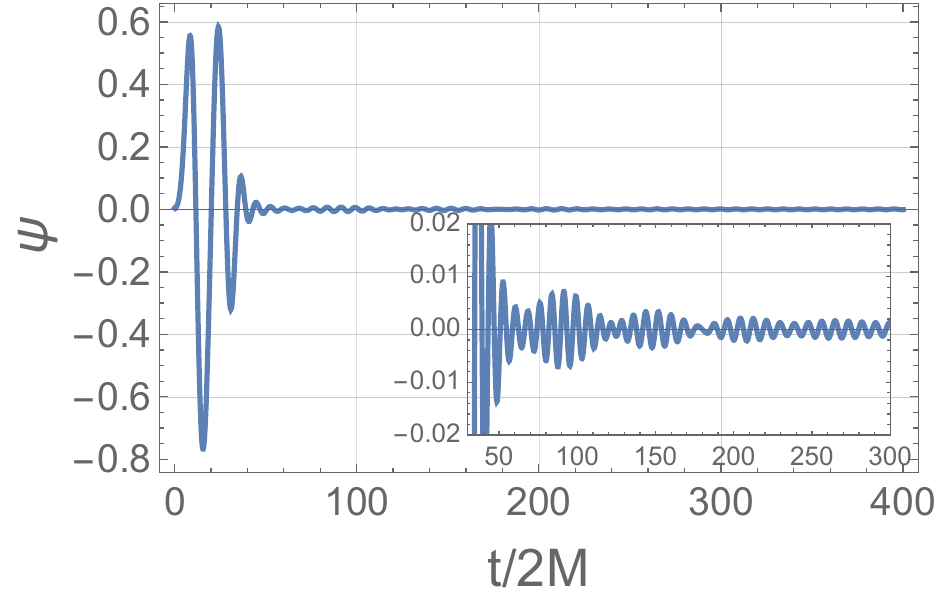}
}
{
\label{fig:b}
\includegraphics[width=0.31\columnwidth]{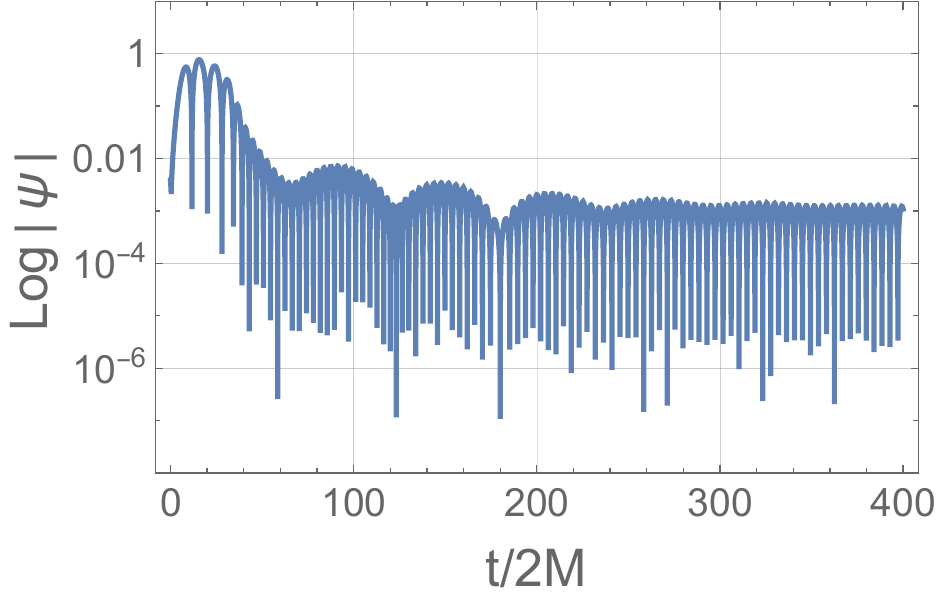}
}
{
\label{fig:b}
\includegraphics[width=0.31\columnwidth]{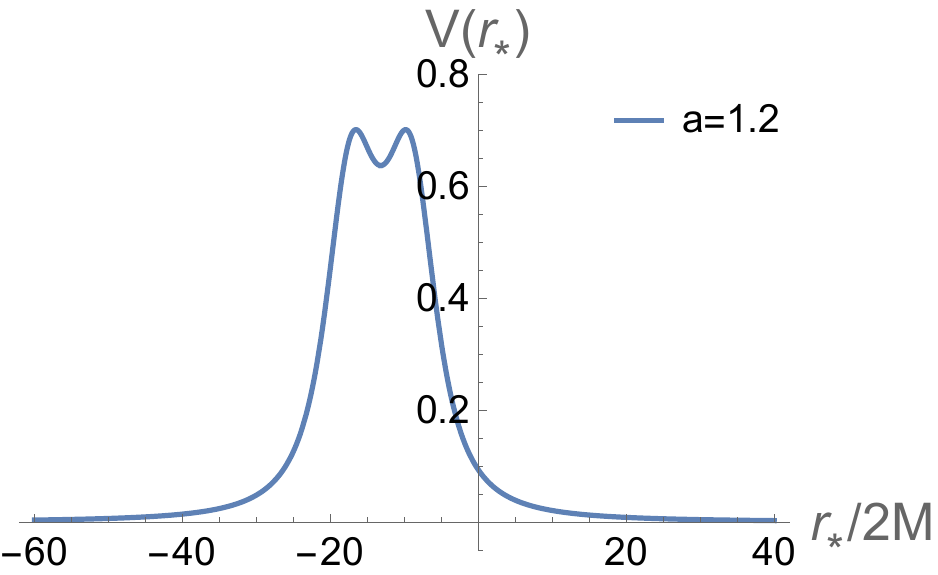}
}
{
\label{fig:b}
\includegraphics[width=0.31\columnwidth]{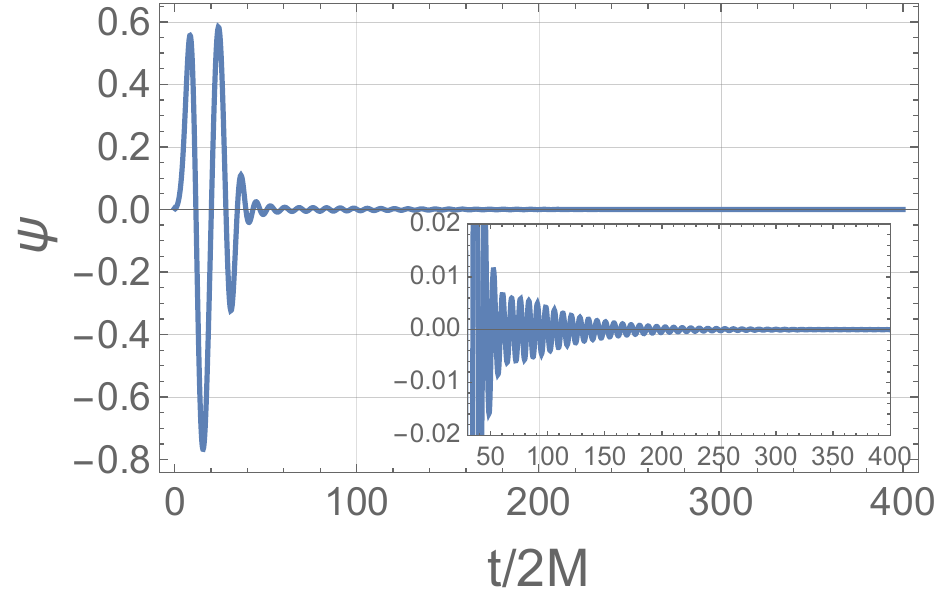}
}
{
\label{fig:b}
\includegraphics[width=0.31\columnwidth]{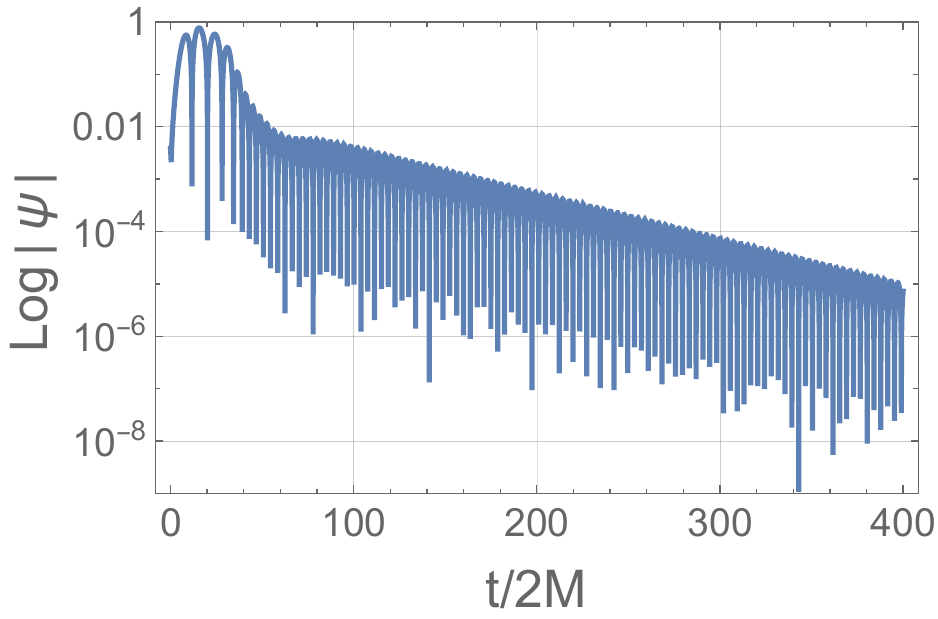}
}
\caption{The effective potentials (left panel), time evolutions and its semilogarithmic plots (right panel) of black holes with the different geometric parameters $a$ in a dark matter halo of M87 under the axial gravitational perturbation. The parameters we used are $M=1/2, \alpha=0.16, \rho_e=6.9 \times 10^6 M_\odot/kpc^3, r_e=91.2kpc$. We have converted these main calculation parameters to the black hole units before plotting.}
\label{f6}
\end{figure*}
\begin{figure*}[t!]
\centering
{
\label{fig:b}
\includegraphics[width=0.45\columnwidth]{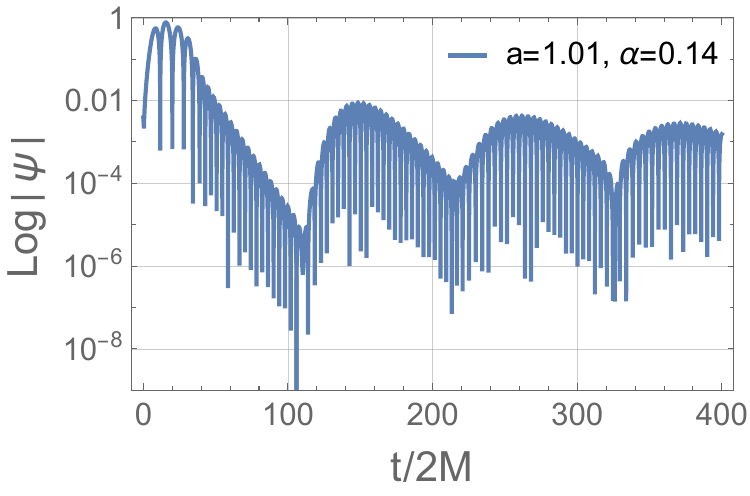}
}
{
\label{fig:b}
\includegraphics[width=0.45\columnwidth]{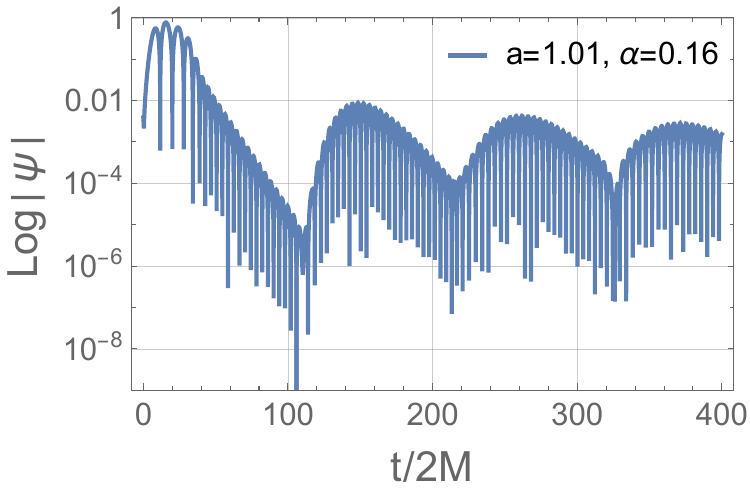}
}
{
\label{fig:b}
\includegraphics[width=0.45\columnwidth]{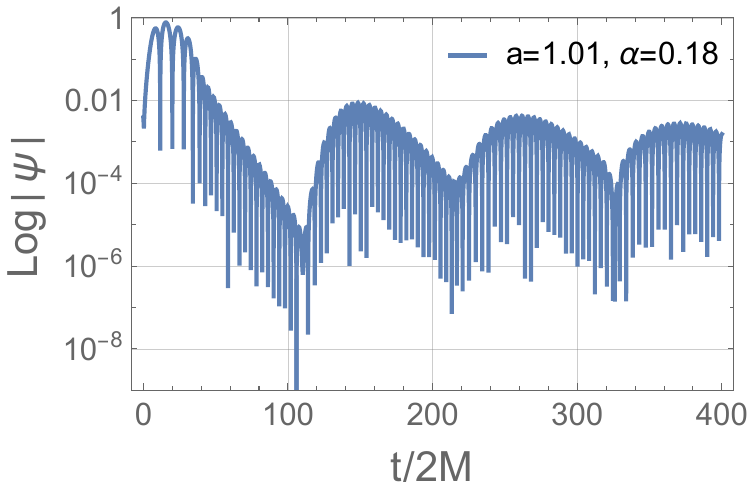}
}
{
\label{fig:b}
\includegraphics[width=0.45\columnwidth]{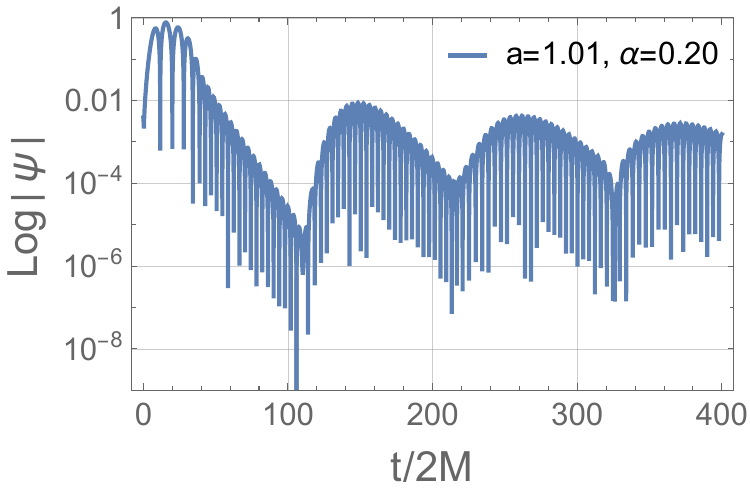}
}
\caption{The semilogarithmic plots of the time evolutions in black holes with the different shape parameter $\alpha$ for the geometric parameter $a=1.01$ in a dark matter halo of M87 under the axial gravitational perturbation. The parameters we used are $M=1/2, a=1.01, \rho_e=6.9 \times 10^6 M_\odot/kpc^3, r_e=91.2kpc$. We have converted these main calculation parameters to the black hole units before plotting.}
\label{f7}
\end{figure*}

\subsection{Quasinormal modes and gravitational wave echoes of black holes in a dark matter halo}
In this subsection, we will use the numerical methods introduced in the previous subsection to calculate the quasinormal mode (QNM) of black holes and study the impacts of dark matter parameters on the QNM in the dark matter halo of M87 under axial gravitational perturbation. Firstly, we use the WKB method to calculate the QNM of the black hole for the geometric parameter $a \leq 2M$ ($M=1/2$) because in this case the effective potential of a black hole only has one maximum. In Table \ref{t1}, we present the fundamental quasinormal modes of a black hole in a dark matter halo of M87 under the different geometric parameter $a$ and shape parameter $\alpha$. Our results show that under the same shape parameter, the real part of the QNM increases with the increasing of the geometric parameter $a$, while the absolute value of the imaginary part decreases with the increasing of the geometric parameter $a$. On the other hand, under the same geometric parameter $a$, both the real part and the absolute values of the imaginary part of the QNM decrease with the increasing of the shape parameter $\alpha$. In particular, with the increasing of the shape parameter $\alpha$, we find that when the shape parameter $\alpha=0.18$, the frequencies of the quasinormal modes of black holes in a dark matter halo are the same as the case of the shape parameter $\alpha=0.20$. However, the QNM frequencies between the two different parameters must be different. In other words, when the geometric parameter $a\leq2M$, with the increasing of the shape parameter $\alpha$, the QNM frequencies of black holes appear to be zero difference. Zero difference of the frequencies is not allowed, so there is a theoretical upper limit on the shape parameter $\alpha$. From this perspective, this upper limit of the shape parameter $\alpha$ when the geometric parameter $a\leq2M$, which is approximately $\alpha<0.18$. With the continuous development of detection technology, we believe these results may provide a new method for the indirect search for dark matter.

\indent On the other hand, when the geometric parameter $a>2M$ ($M=1/2$), the regular black hole transforms into a traversable wormhole, and then the gravitational wave echoes are generated. Here, we are interested in the time evolution of the gravitational wave echoes of a black hole in a dark matter halo of M87 and its frequencies. We use the time domain method to give the time evolution of the gravitational wave echoes, and then use the Prony method to extract the frequencies of the quasinormal modes and the gravitational wave echoes. In Figure \ref{f5}, we show the effective potential and time evolution of black holes in a dark matter halo with the geometric parameters $a=0.9, 1, 1.01$, respectively. Our results show that the gravitational wave echoes appear when the geometric parameter $a>2M$, which corresponds to a double barrier feature of the effective potential of a black hole. When the effective potential has a single barrier structure, there are no gravitational wave echoes. In other words, one of the conditions for the generation of gravitational wave echoes is that the effective potential of the black hole has a double barrier feature. \\
\indent Next, we continue to study and analyze the gravitational wave echoes of the black hole in a dark matter halo of M87. In Figure \ref{f6}, we show the effective potentials, time evolution and its semilogarithmic plots of a black hole under the different geometric parameter $a$, respectively. Our results show that the black hole gravitational wave echoes gradually disappear with the increasing of the geometric parameter $a$. For the case of the geometric parameter $a=1.01$, we can clearly see a series of echoes appearing after the quasinormal mode. Therefore, at the geometric parameter $a=1.01$, the correlation properties of the echoes can be accurately observed. Therefore, as an example, we study the impacts of the dark matter parameters on the gravitational wave echoes when the geometric parameter $a=1.01$. In Figure \ref{f7}, we show the semilogarithmic plots of the time evolution in a black hole with the different shape parameters $\alpha$ for the geometric parameter $a=1.01$ in a dark matter halo of M87 under the axial gravitational perturbation. From Figure \ref{f7}, it is difficult to distinguish the impacts of the shape parameters of dark matter on the time evolutions of a black hole, because their behaves of these images between them are very similar. A reasonable solution would be to extract their frequencies from various stages of time evolution with sufficient accuracy. Therefore, in Table \ref{t2}, we use the Prony method to extract the frequencies of a black hole in the fundamental quasinormal mode and the gravitational wave echoes with the shape parameter $\alpha$ from 0.10 to 0.20, respectively. Our results show that the real parts of the frequencies of both the fundamental quasinormal mode and the gravitational wave echoes decrease basically with the increasing of the shape parameter $\alpha$, while the absolute values of their imaginary parts increase basically with the increasing of the shape parameter $\alpha$. In particular, with the increasing of the shape parameter $\alpha$, we find that when the shape parameter $\alpha=0.18$, the frequencies of the quasinormal modes and the gravitational wave echoes in a dark matter halo are the same as the case of the shape parameter $\alpha=0.20$. In other words, when the geometric parameter $a>2M$, with the increasing of the shape parameter $\alpha$, the frequencies of the quasinormal modes and the gravitational wave echoes appear to be zero difference. This is obviously not allowed, therefore, we give an upper limit of the shape parameter $\alpha$ when the geometric parameter $a>2M$, which is approximately $\alpha<0.18$. This conclusion is consistent with the case that the geometric parameter $a\leq2M$. Variations at the geometric parameter $a$ in this gravitational wave echoes corresponding to significant information about the structure of the horizon. In other words, if the frequencies of these gravitational wave echoes can be detected in the near future, it may be possible to prove the existence of the traversable wormholes.

\newpage
\begin{table}[H]
\setlength{\abovecaptionskip}{0.2cm}
\setlength{\belowcaptionskip}{0.2cm}
\begin{minipage}{0.3\textwidth}
\rotatebox{90}{\begin{minipage}{\textheight} 
\centering
\caption{The fundamental quasinormal mode frequencies of a black hole with different geometric parameter $a$ and shape parameter $\alpha$ in a dark matter halo of M87. The method we used is WKB method and the main calculation parameters are $M=1/2$, $\rho_e=6.9 \times 10^{6}$ $M_{\bigodot}/kpc^{3}$, $r_e=91.2$ $kpc$.We have converted these main calculation parameters to the black hole units before calculating.}
\scalebox{.75}{
\begin{tabular}{ccccc}
\hline  \hline
                                    & $\alpha=0.14$                                                  & $\alpha=0.16$                                                  & $\alpha=0.18$                                                  & $\alpha=0.20$                               \\ \hline
$a=0.0$ & 0.746324129709541 - 0.17843489444755$i$ & 0.746324129709558 - 0.17843489444767$i$ & 0.746324129709583 - 0.17843489444772$i$ & 0.746324129709583 - 0.17843489444772$i$  \\
$a=0.1$ & 0.746738669404371 - 0.17786167792088$i$ & 0.746738669404316 - 0.17786167792065$i$ & 0.746738669404616 - 0.17786167792186$i$ & 0.746738669404616 - 0.17786167792186$i$ \\
$a=0.2$ & 0.748005683784469 - 0.17614828971544$i$ & 0.748005683784769 - 0.17614828971660$i$ & 0.748005683784589 - 0.17614828971586$i$ & 0.748005683784589 - 0.17614828971586$i$  \\
$a=0.3$ & 0.750196429763793 - 0.17331881522164$i$ & 0.750196429763772 - 0.17331881522127$i$ & 0.750196429763687 - 0.17331881522098$i$ & 0.750196429763687 - 0.17331881522098$i$   \\
$a=0.4$ & 0.753431417047884 - 0.16943077482210$i$ & 0.753431417047767 - 0.16943077482168$i$ & 0.753431417047874 - 0.16943077482210$i$ & 0.753431417047874 - 0.16943077482210$i$ \\
$a=0.5$ & 0.757874607005585 - 0.16459620845755$i$ & 0.757874607005599 - 0.16459620845743$i$ & 0.757874607005663 - 0.16459620845786$i$ & 0.757874607005663 - 0.16459620845786$i$  \\
$a=0.6$ & 0.763700279646073 - 0.15898187760955$i$ & 0.763700279645868 - 0.15898187760853$i$ & 0.763700279645807 - 0.15898187760827$i$ & 0.763700279645807 - 0.15898187760827$i$ \\
$a=0.7$ & 0.770989180567407 - 0.15269952669382$i$ & 0.770989180567363 - 0.15269952669360$i$ & 0.770989180567558 - 0.15269952669444$i$ & 0.770989180567558 - 0.15269952669444$i$ \\
$a=0.8$ & 0.779498491760576 - 0.14540400674727$i$ & 0.779498491760671 - 0.14540400674794$i$ & 0.779498491760611 - 0.14540400674757$i$ & 0.779498491760611 - 0.14540400674757$i$  \\
$a=0.9$ & 0.788381626334941 - 0.13561235436104$i$ & 0.788381626334618 - 0.13561235435933$i$ & 0.788381626334400 - 0.13561235435805$i$ & 0.788381626334400 - 0.13561235435805$i$ \\
$a=1.0$ & 0.796399872856131 - 0.12109224424459$i$ & 0.796399872856296 - 0.12109224424554$i$ & 0.796399872856351 - 0.12109224424588$i$ & 0.796399872856351 - 0.12109224424588$i$ \\ \hline \hline
\end{tabular}
\label{t1}}
\end{minipage}
}
\end{minipage}
\hspace{0.2\textwidth} 
\begin{minipage}{0.3\textwidth}
\rotatebox{90}{\begin{minipage}{\textheight} 
\centering
\caption{The fundamental quasinormal mode and gravitational wave echoes frequencies of black holes with different shape parameter $\alpha$ for the geometric parameter $a=1.01$ in a dark matter halo of M87. The method we used is Prony method and the main calculation parameters are $M=1/2, a=1.01$, $\rho_e=6.9 \times 10^{6}$ $M_{\bigodot}/kpc^{3}$, $r_e=91.2$ $kpc$. We have converted these main calculation parameters to the black hole units before calculating.}
\scalebox{0.98}{
\begin{tabular}{cccc}
\hline \hline
              & fundamental quasinormal mode                         & first echo                            & second echo                            \\ \hline
$\alpha=0.10$ & 0.79402110592883 - 0.14030703356915$i$ & 0.79972466361203 - 0.08892365061043$i$ & 0.77346034062292 - 0.05514630691471$i$ \\
$\alpha=0.12$ & 0.79402110592771 - 0.14030703356973$i$ & 0.79972466362508 - 0.08892365060973$i$ & 0.77346034062963 - 0.05514630691546$i$ \\
$\alpha=0.14$ & 0.79402110592716 - 0.14030703357038$i$ & 0.79972466361630 - 0.08892365061085$i$ & 0.77346034064037 - 0.05514630691501$i$ \\
$\alpha=0.16$ & 0.79402110592935 - 0.14030703357010$i$ & 0.79972466361335 - 0.08892365061000$i$ & 0.77346034062732 - 0.05514630691562$i$ \\
$\alpha=0.18$ & 0.79402110592621 - 0.14030703357057$i$ & 0.79972466359582 - 0.08892365061000$i$ & 0.77346034063336 - 0.05514630691460$i$ \\
$\alpha=0.20$ & 0.79402110592621 - 0.14030703357057$i$ & 0.79972466359582 - 0.08892365061000$i$ & 0.77346034063336 - 0.05514630691460$i$ \\ \hline \hline
\end{tabular}
\label{t2}}
\end{minipage}
}
\end{minipage}
\end{table}
\subsection{The detectability of the impacts on the QNM of black holes in a dark matter halo}
In the previous subsection, we studied and analyzed the fundamental quasinormal mode and gravitational wave echoes of the black holes in a dark matter halo of M87, and studied impacts of the shape parameter $\alpha$ of dark matter on their QNM frequencies. Therefore, in this subsection, we are interested in whether these impacts of dark matter on the black hole has the possibility of detection. Under the axial gravitational perturbation, the ringdown signal of gravitational waves can be written as follows \cite{Berti:2005ys},
\begin{equation}
h(t)=h_+(t)+ih_\times (t)=\frac{M}{r}\sum_{lmn}A_{lmn}e^{i(f_{lmn}t+\phi_{lmn})e^{-t/\tau_{lmn}}}S_{lmn},
\end{equation}
where, $M$ stands for the redshift mass of a black hole, $r$ stands for the luminosity distance to the source, $A_{lmn}$ stands for the amplitude of the fundamental quasinormal mode,  $\phi_{lmn}$ stands for the phase coefficient and $S_{lmn}$ stands for the spheroidal harmonics of the spin weight 2. For the black hole in a dark matter halo, the frequency of the fundamental quasinormal mode $f_{lmn}$ and the  damping time $\tau_{lmn}$  are given by
\begin{equation}
2\pi f_{lmn}=\text{Re}(\omega_{lmn}), \quad \tau_{lmn}=-1/\text{Im}(\omega_{lmn})
\end{equation}
where, $\omega_{lmn}$ stands for the fundamental quasinormal mode of the black hole in a dark matter halo of M87 under axial gravitational perturbation with $(l,m,n)=(2,0,0)$. Based on the Ref. \cite{Zhang:2022roh}, the frequency and damping time of the black hole in a dark matter halo of M87 can be given by
\begin{equation}
f_{lmn}=f_{lmn}^{\text{BBS}}(1+\delta f_{lmn}),\quad \tau_{lmn}=\tau_{lmn}^{\text{BBS}}(1+\delta \tau_{lmn}),
\end{equation}
where, $f_{lmn}^{\text{BBS}}, \tau_{lmn}^{\text{BBS}}$ are the frequency and damping time of the black bounce spacetime.
$\delta f_{lmn},\delta \tau_{lmn}$ are the relative deviations to the black bounce spacetime. In Table. \ref{t3}, our results show that the relative deviation is approximately $10^{-14} \sim  10^{-13}$ for the frequency and $10^{-12} \sim  10^{-11}$ for the damping time. Although it is currently unlikely to detect such relative deviations, with the continuous development of detecting technology, we believe that we will be able to detect these signals with lower relative deviations in the near future, which may also provide an effective method for the indirect search for dark matter.
\begin{table}[h!]
\setlength{\abovecaptionskip}{0.2cm}
\setlength{\belowcaptionskip}{0.2cm}
\centering
\caption{The relative deviations of the frequency $f_{200}$ and damping time $\tau_{200}$ in the traversable wormhole with different geometric parameters for the shape parameter under a dark matter halo of M87. The method we used is Prony method and the main calculation parameters are $M=1/2, \alpha=0.16$, $\rho_e=6.9 \times 10^{6}$ $M_{\bigodot}/kpc^{3}$, $r_e=91.2$ $kpc$. We have converted these main calculation parameters to the black hole units before calculating.}
\scalebox{0.72}{
\begin{tabular}{ccccccccc}
\hline \hline
                    & $a=0.00 $                  & $a=0.20 $                           & $a=0.40 $                  & $a=0.60  $                 & $a=0.80$                   & $a=1.00$                   & $a=1.01$                   & $a=1.10 $                  \\ \hline
$\delta f_{200}$      & 2.04 $\times 10^{-13}$ & 2.41 $\times 10^{-13}$ & 1.42 $\times 10^{-13}$ & 1.05 $\times 10^{-13}$ & 7.72 $\times 10^{-14}$ & 6.96 $\times 10^{-14}$ & 2.66 $\times 10^{-13}$ & 8.95 $\times 10^{-13}$ \\
$\delta \tau_{200}$ & 3.60 $\times 10^{-12}$ & 4.15 $\times 10^{-12}$ & 2.47 $\times 10^{-12}$ & 3.53 $\times 10^{-12}$ & 2.53 $\times 10^{-12}$ & 2.79 $\times 10^{-12}$ & 1.24 $\times 10^{-11}$  & 5.92 $\times 10^{-11}$  \\ \hline \hline
\end{tabular}
\label{t3}}
\end{table}

\section{Summary and discussion}\label{s6}
In this work, combining the mass model of M87 and the Einasto profile of dark matter halo, we construct one formal solution of black holes in a dark matter halo through relativistic analysis. We then test this solution of black holes under the axial gravitational perturbation and calculate their quasinormal modes (QNM). Except the QNM, we also discover the existence of the gravitational wave echoes. In addition, we also study the impacts of shape parameters $\alpha$ of the Einasto profile on the QNM and gravitational wave echoes. Finally, we analyze the possibility of astronomical detectors detecting the impacts of dark matter parameter on QNM. The main conclusions are as follows:

(1) Combining the mass model of M87 and the Einasto profile of a dark matter halo, we construct one formal solution of black holes from Einstein field equations and now this solution is given by Eq. (\ref{e37}), which includes the regular black hole for the geometric parameter $a < 2M$ and the wormhole for the geometric parameter $a\geq2M$. This solution corresponds to black holes with the astronomical environment full of dark matter. If the dark matter is absent, this black hole solution degenerates into the solution of the black bounce spacetime.

(2) We test this solution of black holes under the axial gravitational perturbation and calculate their effective potentials and QNMs. In this solution, when the geometric parameter $a \leq 2M$, the maximum of the effective potential decreases with the increasing of the geometric parameter $a$. Then, we use the WKB method to extract the QNM frequencies of the black hole. The real part of the QNM frequencies in the black hole increases with the increasing of the geometric parameter $a$, while the absolute value of the imaginary part decreases. Besides, the impacts of shape parameter $\alpha$ in the Einasto profile on QNM have been studied. When the geometric parameter $a$ is fixed, the real part and the absolute value of imaginary part of QNM frequencies decrease with the increasing of the shape parameter $\alpha$. On the other hand, with the increasing of the shape parameter $\alpha$, when the shape parameter $\alpha=0.18$, the frequencies of the quasinormal modes of black holes in a dark matter halo are the same as the case of the shape parameter $\alpha=0.20$. In other words, with the increasing of the shape parameter $\alpha$, the frequencies of the quasinormal modes of black holes appear to be zero difference. Zero difference of the frequencies is not allowed, so there is a theoretical upper limit on the shape parameter $\alpha$. From this perspective, this upper limit of the shape parameter $\alpha$when the geometric parameter $a\leq2M$, which is approximately $\alpha<0.18$. These results may provide some assistance for the indirect search for dark matter.

(3) In this solution, when the geometric parameter $a>2M$ $(M=1/2)$, the maximum of the effective potential increases with the increasing of the geometric parameter $a$, and the effective potential has an obvious double barrier feature. Then with the geometric parameter $a$ continually increasing, the potential well between the double barrier rises and narrows, eventually returning to a single potential barrier. From the time evolution of the QNM, when the geometric parameter $a>1$, after the QNM is a series of the gravitational wave echoes. In other words, one of the distinctive features of the gravitational wave echoes is the double barrier feature. Finally, we extract the frequencies of the QNM and the gravitational wave echoes using the Prony method. When the geometric parameter $a=1.01$, the real part of the QNM frequencies decreases with the increasing of the shape parameter $\alpha$, while the absolute value of the imaginary part increases. For the frequencies of the gravitational wave echoes, the real parts of the frequencies of the first echo and the second echo basically decrease with the increasing of the shape parameter $\alpha$, while the absolute value of the imaginary part increases. On the other hand, when the geometric parameter $a>2M$, with the increasing of the shape parameter $\alpha$, the frequencies of the quasinormal modes and the gravitational wave echoes appear to be zero difference. This is obviously not allowed, therefore, we give an upper limit of the shape parameter $\alpha$ when the geometric parameter $a>2M$, which is approximately $\alpha<0.18$. This conclusion is consistent with the case that the geometric parameter $a\leq2M$. Variations at the geometric parameter $a$ between the QNM and the gravitational wave echoes can provide significant information about the structure of the horizon. In other words, if the frequencies of these gravitational wave echoes can be detected in the near future, it may be possible to prove the existence of the traversable wormholes.

(4) Finally, we analyze the possibility of astronomical detectors detecting the impacts of dark matter on the QNM in this black hole. When the shape parameter $\alpha=0.16$, the relative deviation of a black hole in a dark matter halo is approximately $f_{200}=10^{-14} \sim 10^{-13}$ for the frequency and $\tau_{200}=10^{-12} \sim 10 ^{-11}$ for the damping time. Although it is currently unlikely to detect such relative deviations, with the continuous development of observation technology, we believe that these small changes will be able to be detected in the near future, which may provide some help for the search for dark matter.

The last but not the least, these results above are based on the black holes in a dark matter halo with Einasto profile. On the other hand, P. Gondolo et al. show that the density of dark matter at the galactic center may be enhanced Because of the presence of a black hole, leading to a ``spike'' phenomenon \cite{Gondolo:1999ef}. In a further step of work, we will consider whether such spikes can be studied through gravitational wave echoes, in order to provide more possibilities for indirect detection of dark matter.

\acknowledgments
We would like to acknowledge the anonymous referee for the constructive report, which significantly improved this paper. This research was funded by the National Natural Science Foundation of China (Grant No.12265007).

\bibliographystyle{epjc}
\bibliography{epjcexample}

\providecommand{\href}[2]{#2}\begingroup\raggedright\begin{thebibliography}{10}

\bibitem{LIGO2}
{\bf LIGO Scientific, Virgo} Collaboration, B.~P. Abbott et~al., {\it
  {Observation of Gravitational Waves from a Binary Black Hole Merger}},  {\em
  Phys. Rev. Lett.} {\bf 116} (2016), no.~6 061102,
  [\href{http://arxiv.org/abs/1602.03837}{{\tt arXiv:1602.03837}}].

\bibitem{LIGO1}
{\bf LIGO Scientific, Virgo} Collaboration, B.~P. Abbott et~al., {\it
  {GW151226: Observation of Gravitational Waves from a 22-Solar-Mass Binary
  Black Hole Coalescence}},  {\em Phys. Rev. Lett.} {\bf 116} (2016), no.~24
  241103, [\href{http://arxiv.org/abs/1606.04855}{{\tt arXiv:1606.04855}}].

\bibitem{LIGO3}
{\bf LIGO Scientific, Virgo} Collaboration, B.~P. Abbott et~al., {\it
  {GW170814: A Three-Detector Observation of Gravitational Waves from a Binary
  Black Hole Coalescence}},  {\em Phys. Rev. Lett.} {\bf 119} (2017), no.~14
  141101, [\href{http://arxiv.org/abs/1709.09660}{{\tt arXiv:1709.09660}}].

\bibitem{LIGO4}
{\bf LIGO Scientific, Virgo} Collaboration, B.~P. Abbott et~al., {\it
  {GW170817: Observation of Gravitational Waves from a Binary Neutron Star
  Inspiral}},  {\em Phys. Rev. Lett.} {\bf 119} (2017), no.~16 161101,
  [\href{http://arxiv.org/abs/1710.05832}{{\tt arXiv:1710.05832}}].

\bibitem{EventHorizonTelescope:2019dse}
{\bf Event Horizon Telescope} Collaboration, K.~Akiyama et~al., {\it {First M87
  Event Horizon Telescope Results. I. The Shadow of the Supermassive Black
  Hole}},  {\em Astrophys. J. Lett.} {\bf 875} (2019) L1,
  [\href{http://arxiv.org/abs/1906.11238}{{\tt arXiv:1906.11238}}].

\bibitem{EventHorizonTelescope:2022apq}
{\bf Event Horizon Telescope} Collaboration, K.~Akiyama et~al., {\it {First
  Sagittarius A* Event Horizon Telescope Results. II. EHT and Multiwavelength
  Observations, Data Processing, and Calibration}},  {\em Astrophys. J. Lett.}
  {\bf 930} (2022), no.~2 L13.

\bibitem{LIGOScientific:2016aoc}
{\bf LIGO Scientific, Virgo} Collaboration, B.~P. Abbott et~al., {\it
  {Observation of Gravitational Waves from a Binary Black Hole Merger}},  {\em
  Phys. Rev. Lett.} {\bf 116} (2016), no.~6 061102,
  [\href{http://arxiv.org/abs/1602.03837}{{\tt arXiv:1602.03837}}].

\bibitem{Konoplya:2011qq}
R.~A. Konoplya and A.~Zhidenko, {\it {Quasinormal modes of black holes: From
  astrophysics to string theory}},  {\em Rev. Mod. Phys.} {\bf 83} (2011)
  793--836, [\href{http://arxiv.org/abs/1102.4014}{{\tt arXiv:1102.4014}}].

\bibitem{Rahman:2021kwb}
M.~Rahman and A.~Bhattacharyya, {\it {Ringdown of charged compact objects using
  membrane paradigm}},  {\em Phys. Rev. D} {\bf 104} (2021), no.~4 044045,
  [\href{http://arxiv.org/abs/2104.00074}{{\tt arXiv:2104.00074}}].

\bibitem{Simpson:2018tsi}
A.~Simpson and M.~Visser, {\it {Black-bounce to traversable wormhole}},  {\em
  JCAP} {\bf 02} (2019) 042, [\href{http://arxiv.org/abs/1812.07114}{{\tt
  arXiv:1812.07114}}].

\bibitem{Simpson:2019cer}
A.~Simpson, P.~Martin-Moruno, and M.~Visser, {\it {Vaidya spacetimes,
  black-bounces, and traversable wormholes}},  {\em Class. Quant. Grav.} {\bf
  36} (2019), no.~14 145007, [\href{http://arxiv.org/abs/1902.04232}{{\tt
  arXiv:1902.04232}}].

\bibitem{Churilova:2019cyt}
M.~S. Churilova and Z.~Stuchlik, {\it {Ringing of the regular
  black-hole/wormhole transition}},  {\em Class. Quant. Grav.} {\bf 37} (2020),
  no.~7 075014, [\href{http://arxiv.org/abs/1911.11823}{{\tt
  arXiv:1911.11823}}].

\bibitem{Pal:2022cxb}
K.~Pal, K.~Pal, P.~Roy, and T.~Sarkar, {\it {Regularizing the JNW and JMN naked
  singularities}},  {\em Eur. Phys. J. C} {\bf 83} (2023), no.~5 397,
  [\href{http://arxiv.org/abs/2206.11764}{{\tt arXiv:2206.11764}}].

\bibitem{Mazur:2001fv}
P.~O. Mazur and E.~Mottola, {\it {Gravitational Condensate Stars: An
  Alternative to Black Holes}},  {\em Universe} {\bf 9} (2023), no.~2 88,
  [\href{http://arxiv.org/abs/gr-qc/0109035}{{\tt gr-qc/0109035}}].

\bibitem{Almheiri:2012rt}
A.~Almheiri, D.~Marolf, J.~Polchinski, and J.~Sully, {\it {Black Holes:
  Complementarity or Firewalls?}},  {\em JHEP} {\bf 02} (2013) 062,
  [\href{http://arxiv.org/abs/1207.3123}{{\tt arXiv:1207.3123}}].

\bibitem{Vicente:2018mxl}
R.~Vicente, V.~Cardoso, and J.~C. Lopes, {\it {Penrose process, superradiance,
  and ergoregion instabilities}},  {\em Phys. Rev. D} {\bf 97} (2018), no.~8
  084032, [\href{http://arxiv.org/abs/1803.08060}{{\tt arXiv:1803.08060}}].

\bibitem{Bueno:2017hyj}
P.~Bueno, P.~A. Cano, F.~Goelen, T.~Hertog, and B.~Vercnocke, {\it {Echoes of
  Kerr-like wormholes}},  {\em Phys. Rev. D} {\bf 97} (2018), no.~2 024040,
  [\href{http://arxiv.org/abs/1711.00391}{{\tt arXiv:1711.00391}}].

\bibitem{Maselli:2017cmm}
A.~Maselli, P.~Pani, V.~Cardoso, T.~Abdelsalhin, L.~Gualtieri, and V.~Ferrari,
  {\it {Probing Planckian corrections at the horizon scale with LISA
  binaries}},  {\em Phys. Rev. Lett.} {\bf 120} (2018), no.~8 081101,
  [\href{http://arxiv.org/abs/1703.10612}{{\tt arXiv:1703.10612}}].

\bibitem{Cardoso:2016oxy}
V.~Cardoso, S.~Hopper, C.~F.~B. Macedo, C.~Palenzuela, and P.~Pani, {\it
  {Gravitational-wave signatures of exotic compact objects and of quantum
  corrections at the horizon scale}},  {\em Phys. Rev. D} {\bf 94} (2016),
  no.~8 084031, [\href{http://arxiv.org/abs/1608.08637}{{\tt
  arXiv:1608.08637}}].

\bibitem{Cardoso:2016rao}
V.~Cardoso, E.~Franzin, and P.~Pani, {\it {Is the gravitational-wave ringdown a
  probe of the event horizon?}},  {\em Phys. Rev. Lett.} {\bf 116} (2016),
  no.~17 171101, [\href{http://arxiv.org/abs/1602.07309}{{\tt
  arXiv:1602.07309}}]. [Erratum: Phys.Rev.Lett. 117, 089902 (2016)].

\bibitem{Dong:2020odp}
R.~Dong and D.~Stojkovic, {\it {Gravitational wave echoes from black holes in
  massive gravity}},  {\em Phys. Rev. D} {\bf 103} (2021), no.~2 024058,
  [\href{http://arxiv.org/abs/2011.04032}{{\tt arXiv:2011.04032}}].

\bibitem{Yang:2021cvh}
Y.~Yang, D.~Liu, Z.~Xu, Y.~Xing, S.~Wu, and Z.-W. Long, {\it {Echoes of novel
  black-bounce spacetimes}},  {\em Phys. Rev. D} {\bf 104} (2021), no.~10
  104021, [\href{http://arxiv.org/abs/2107.06554}{{\tt arXiv:2107.06554}}].

\bibitem{PhysRevD.49.883}
C.~Gundlach, R.~H. Price, and J.~Pullin, {\it Late-time behavior of stellar
  collapse and explosions. i. linearized perturbations},  {\em Phys. Rev. D}
  {\bf 49} (Jan, 1994) 883--889.

\bibitem{Schutz:1985km}
B.~F. Schutz and C.~M. Will, {\it {BLACK HOLE NORMAL MODES: A SEMIANALYTIC
  APPROACH}},  {\em Astrophys. J. Lett.} {\bf 291} (1985) L33--L36.

\bibitem{PhysRevD.35.3621}
S.~Iyer and C.~M. Will, {\it Black-hole normal modes: A wkb approach. i.
  foundations and application of a higher-order wkb analysis of
  potential-barrier scattering},  {\em Phys. Rev. D} {\bf 35} (Jun, 1987)
  3621--3631.

\bibitem{Poschl:1933zz}
G.~P{\"o}schl and E.~Teller, {\it {Bemerkungen zur Quantenmechanik des
  anharmonischen Oszillators}},  {\em Z. Phys.} {\bf 83} (1933) 143--151.

\bibitem{Ferrari:1984zz}
V.~Ferrari and B.~Mashhoon, {\it {New approach to the quasinormal modes of a
  black hole}},  {\em Phys. Rev. D} {\bf 30} (1984) 295--304.

\bibitem{Churilova:2021nnc}
M.~S. Churilova, R.~A. Konoplya, and A.~Zhidenko, {\it {Analytic formula for
  quasinormal modes in the near-extreme Kerr-Newman\textendash{}de Sitter
  spacetime governed by a non-P\"oschl-Teller potential}},  {\em Phys. Rev. D}
  {\bf 105} (2022), no.~8 084003, [\href{http://arxiv.org/abs/2108.04858}{{\tt
  arXiv:2108.04858}}].

\bibitem{Leaver:1985ax}
E.~W. Leaver, {\it {An Analytic representation for the quasi normal modes of
  Kerr black holes}},  {\em Proc. Roy. Soc. Lond. A} {\bf 402} (1985) 285--298.

\bibitem{Dolan:2007mj}
S.~R. Dolan, {\it {Instability of the massive Klein-Gordon field on the Kerr
  spacetime}},  {\em Phys. Rev. D} {\bf 76} (2007) 084001,
  [\href{http://arxiv.org/abs/0705.2880}{{\tt arXiv:0705.2880}}].

\bibitem{Panotopoulos:2017hns}
G.~Panotopoulos and A.~Rinc\'on, {\it {Quasinormal modes of black holes in
  Einstein-power-Maxwell theory}},  {\em Int. J. Mod. Phys. D} {\bf 27} (2017),
  no.~03 1850034, [\href{http://arxiv.org/abs/1711.04146}{{\tt
  arXiv:1711.04146}}].

\bibitem{Rincon:2018sgd}
A.~Rinc\'on and G.~Panotopoulos, {\it {Quasinormal modes of scale dependent
  black holes in ( 1+2 )-dimensional Einstein-power-Maxwell theory}},  {\em
  Phys. Rev. D} {\bf 97} (2018), no.~2 024027,
  [\href{http://arxiv.org/abs/1801.03248}{{\tt arXiv:1801.03248}}].

\bibitem{Panotopoulos:2019qjk}
G.~Panotopoulos and A.~Rinc\'on, {\it {Quasinormal modes of regular black holes
  with non linear-Electrodynamical sources}},  {\em Eur. Phys. J. Plus} {\bf
  134} (2019), no.~6 300, [\href{http://arxiv.org/abs/1904.10847}{{\tt
  arXiv:1904.10847}}].

\bibitem{Daghigh:2011ty}
R.~G. Daghigh and M.~D. Green, {\it {Validity of the WKB Approximation in
  Calculating the Asymptotic Quasinormal Modes of Black Holes}},  {\em Phys.
  Rev. D} {\bf 85} (2012) 127501, [\href{http://arxiv.org/abs/1112.5397}{{\tt
  arXiv:1112.5397}}].

\bibitem{Campos:2021sff}
J.~A.~V. Campos, M.~A. Anacleto, F.~A. Brito, and E.~Passos, {\it {Quasinormal
  modes and shadow of noncommutative black hole}},  {\em Sci. Rep.} {\bf 12}
  (2022), no.~1 8516, [\href{http://arxiv.org/abs/2103.10659}{{\tt
  arXiv:2103.10659}}].

\bibitem{Matyjasek:2019eeu}
J.~Matyjasek and M.~Telecka, {\it {Quasinormal modes of black holes. II. Pad\'e
  summation of the higher-order WKB terms}},  {\em Phys. Rev. D} {\bf 100}
  (2019), no.~12 124006, [\href{http://arxiv.org/abs/1908.09389}{{\tt
  arXiv:1908.09389}}].

\bibitem{Navarro:1995iw}
J.~F. Navarro, C.~S. Frenk, and S.~D.~M. White, {\it {The Structure of cold
  dark matter halos}},  {\em Astrophys. J.} {\bf 462} (1996) 563--575,
  [\href{http://arxiv.org/abs/astro-ph/9508025}{{\tt astro-ph/9508025}}].

\bibitem{deBlok:2009sp}
W.~J.~G. de~Blok, {\it {The Core-Cusp Problem}},  {\em Adv. Astron.} {\bf 2010}
  (2010) 789293, [\href{http://arxiv.org/abs/0910.3538}{{\tt
  arXiv:0910.3538}}].

\bibitem{Gondolo:1999ef}
P.~Gondolo and J.~Silk, {\it {Dark matter annihilation at the galactic
  center}},  {\em Phys. Rev. Lett.} {\bf 83} (1999) 1719--1722,
  [\href{http://arxiv.org/abs/astro-ph/9906391}{{\tt astro-ph/9906391}}].

\bibitem{Xu:2017bpz}
Z.~Xu, J.~Wang, and X.~Hou, {\it {Kerr\textendash{}anti-de Sitter/de Sitter
  black hole in perfect fluid dark matter background}},  {\em Class. Quant.
  Grav.} {\bf 35} (2018), no.~11 115003,
  [\href{http://arxiv.org/abs/1711.04538}{{\tt arXiv:1711.04538}}].

\bibitem{Xu:2017vse}
Z.~Xu, X.~Hou, X.~Gong, and J.~Wang, {\it {Kerr\textendash{}Newman-AdS black
  hole surrounded by perfect fluid matter in Rastall gravity}},  {\em Eur.
  Phys. J. C} {\bf 78} (2018), no.~6 513,
  [\href{http://arxiv.org/abs/1711.04542}{{\tt arXiv:1711.04542}}].

\bibitem{Xu_2018}
Z.~Xu, X.~Hou, X.~Gong, and J.~Wang, {\it Black hole space-time in dark matter
  halo},  {\em Journal of Cosmology and Astroparticle Physics} {\bf 2018} (sep,
  2018) 038--038.

\bibitem{Xu:2021dkv}
Z.~Xu, J.~Wang, and M.~Tang, {\it {Deformed black hole immersed in dark matter
  spike}},  {\em JCAP} {\bf 09} (2021) 007,
  [\href{http://arxiv.org/abs/2104.13158}{{\tt arXiv:2104.13158}}].

\bibitem{Liu:2023oab}
D.~Liu, Y.~Yang, Z.~Xu, and Z.-W. Long, {\it {Modeling the black holes
  surrounded by a dark matter halo in the galactic center of M87}},  {\em Eur.
  Phys. J. C} {\bf 84} (2024), no.~2 136,
  [\href{http://arxiv.org/abs/2307.13553}{{\tt arXiv:2307.13553}}].

\bibitem{Liu:2022ygf}
D.~Liu, Y.~Yang, A.~\"Ovg\"un, Z.-W. Long, and Z.~Xu, {\it {Gravitational
  ringing and superradiant instabilities of the Kerr-like black holes in a dark
  matter halo}},  {\em Eur. Phys. J. C} {\bf 83} (2023), no.~7 565,
  [\href{http://arxiv.org/abs/2204.11563}{{\tt arXiv:2204.11563}}].

\bibitem{Yang:2023tip}
Y.~Yang, D.~Liu, A.~\"Ovg\"un, G.~Lambiase, and Z.-W. Long, {\it {Black hole
  surrounded by the pseudo-isothermal dark matter halo}},  {\em Eur. Phys. J.
  C} {\bf 84} (2024), no.~1 63, [\href{http://arxiv.org/abs/2308.05544}{{\tt
  arXiv:2308.05544}}].

\bibitem{Liu:2024xcd}
D.~Liu, Y.~Yang, and Z.-W. Long, {\it {Probing black holes in a dark matter
  spike of M87 using quasinormal modes}},  {\em Eur. Phys. J. C} {\bf 84}
  (2024), no.~7 731, [\href{http://arxiv.org/abs/2401.09182}{{\tt
  arXiv:2401.09182}}].

\bibitem{Zhang:2021bdr}
C.~Zhang, T.~Zhu, and A.~Wang, {\it {Gravitational axial perturbations of
  Schwarzschild-like black holes in dark matter halos}},  {\em Phys. Rev. D}
  {\bf 104} (2021), no.~12 124082, [\href{http://arxiv.org/abs/2111.04966}{{\tt
  arXiv:2111.04966}}].

\bibitem{Zhang:2022roh}
C.~Zhang, T.~Zhu, X.~Fang, and A.~Wang, {\it {Imprints of dark matter on
  gravitational ringing of supermassive black holes}},  {\em Phys. Dark Univ.}
  {\bf 37} (2022) 101078, [\href{http://arxiv.org/abs/2201.11352}{{\tt
  arXiv:2201.11352}}].

\bibitem{Jusufi:2019nrn}
K.~Jusufi, M.~Jamil, P.~Salucci, T.~Zhu, and S.~Haroon, {\it {Black Hole
  Surrounded by a Dark Matter Halo in the M87 Galactic Center and its
  Identification with Shadow Images}},  {\em Phys. Rev. D} {\bf 100} (2019),
  no.~4 044012, [\href{http://arxiv.org/abs/1905.11803}{{\tt
  arXiv:1905.11803}}].

\bibitem{Vagnozzi:2022moj}
S.~Vagnozzi et~al., {\it {Horizon-scale tests of gravity theories and
  fundamental physics from the Event Horizon Telescope image of Sagittarius
  A}},  {\em Class. Quant. Grav.} {\bf 40} (2023), no.~16 165007,
  [\href{http://arxiv.org/abs/2205.07787}{{\tt arXiv:2205.07787}}].

\bibitem{Cardoso:2022whc}
V.~Cardoso, K.~Destounis, F.~Duque, R.~Panosso~Macedo, and A.~Maselli, {\it
  {Gravitational Waves from Extreme-Mass-Ratio Systems in Astrophysical
  Environments}},  {\em Phys. Rev. Lett.} {\bf 129} (2022), no.~24 241103,
  [\href{http://arxiv.org/abs/2210.01133}{{\tt arXiv:2210.01133}}].

\bibitem{Duque:2023cac}
F.~Duque, C.~F.~B. Macedo, R.~Vicente, and V.~Cardoso, {\it {Axion Weak Leaks:
  extreme mass-ratio inspirals in ultra-light dark matter}},
  \href{http://arxiv.org/abs/2312.06767}{{\tt arXiv:2312.06767}}.

\bibitem{Speeney:2024mas}
N.~Speeney, E.~Berti, V.~Cardoso, and A.~Maselli, {\it {Black holes surrounded
  by generic matter distributions: polar perturbations and energy flux}},
  \href{http://arxiv.org/abs/2401.00932}{{\tt arXiv:2401.00932}}.

\bibitem{Cardoso:2021wlq}
V.~Cardoso, K.~Destounis, F.~Duque, R.~P. Macedo, and A.~Maselli, {\it {Black
  holes in galaxies: Environmental impact on gravitational-wave generation and
  propagation}},  {\em Phys. Rev. D} {\bf 105} (2022), no.~6 L061501,
  [\href{http://arxiv.org/abs/2109.00005}{{\tt arXiv:2109.00005}}].

\bibitem{Konoplya:2021ube}
R.~A. Konoplya, {\it {Black holes in galactic centers: Quasinormal ringing,
  grey-body factors and Unruh temperature}},  {\em Phys. Lett. B} {\bf 823}
  (2021) 136734, [\href{http://arxiv.org/abs/2109.01640}{{\tt
  arXiv:2109.01640}}].

\bibitem{Konoplya:2022hbl}
R.~A. Konoplya and A.~Zhidenko, {\it {Solutions of the Einstein Equations for a
  Black Hole Surrounded by a Galactic Halo}},  {\em Astrophys. J.} {\bf 933}
  (2022), no.~2 166, [\href{http://arxiv.org/abs/2202.02205}{{\tt
  arXiv:2202.02205}}].

\bibitem{Shen:2023erj}
Z.~Shen, A.~Wang, Y.~Gong, and S.~Yin, {\it {Analytical models of supermassive
  black holes in galaxies surrounded by dark matter halos}},
  \href{http://arxiv.org/abs/2311.12259}{{\tt arXiv:2311.12259}}.

\bibitem{Figueiredo:2023gas}
E.~Figueiredo, A.~Maselli, and V.~Cardoso, {\it {Black holes surrounded by
  generic dark matter profiles: Appearance and gravitational-wave emission}},
  {\em Phys. Rev. D} {\bf 107} (2023), no.~10 104033,
  [\href{http://arxiv.org/abs/2303.08183}{{\tt arXiv:2303.08183}}].

\bibitem{Xavier:2023exm}
S.~V. M. C.~B. Xavier, H.~C.~D. Lima, Junior., and L.~C.~B. Crispino, {\it
  {Shadows of black holes with dark matter halo}},  {\em Phys. Rev. D} {\bf
  107} (2023), no.~6 064040, [\href{http://arxiv.org/abs/2303.17666}{{\tt
  arXiv:2303.17666}}].

\bibitem{Hayward:2005gi}
S.~A. Hayward, {\it {Formation and evaporation of regular black holes}},  {\em
  Phys. Rev. Lett.} {\bf 96} (2006) 031103,
  [\href{http://arxiv.org/abs/gr-qc/0506126}{{\tt gr-qc/0506126}}].

\bibitem{Ayon-Beato:1999kuh}
E.~Ayon-Beato and A.~Garcia, {\it {New regular black hole solution from
  nonlinear electrodynamics}},  {\em Phys. Lett. B} {\bf 464} (1999) 25,
  [\href{http://arxiv.org/abs/hep-th/9911174}{{\tt hep-th/9911174}}].

\bibitem{Abdujabbarov:2016hnw}
A.~Abdujabbarov, M.~Amir, B.~Ahmedov, and S.~G. Ghosh, {\it {Shadow of rotating
  regular black holes}},  {\em Phys. Rev. D} {\bf 93} (2016), no.~10 104004,
  [\href{http://arxiv.org/abs/1604.03809}{{\tt arXiv:1604.03809}}].

\bibitem{Yang:2023agi}
Y.~Yang, D.~Liu, A.~\"Ovg\"un, G.~Lambiase, and Z.-W. Long, {\it {Rotating
  black hole mimicker surrounded by the string cloud}},  {\em Phys. Rev. D}
  {\bf 109} (2024), no.~2 024002, [\href{http://arxiv.org/abs/2307.09344}{{\tt
  arXiv:2307.09344}}].

\bibitem{Balart:2023odm}
L.~Balart, G.~Panotopoulos, and A.~Rinc\'on, {\it {Regular Charged Black Holes,
  Energy Conditions, and Quasinormal Modes}},  {\em Fortsch. Phys.} {\bf 71}
  (2023), no.~12 2300075, [\href{http://arxiv.org/abs/2309.01910}{{\tt
  arXiv:2309.01910}}].

\bibitem{Konoplya:2024hfg}
R.~A. Konoplya and A.~Zhidenko, {\it {Infinite tower of higher-curvature
  corrections: Quasinormal modes and late-time behavior of D-dimensional
  regular black holes}},  {\em Phys. Rev. D} {\bf 109} (2024), no.~10 104005,
  [\href{http://arxiv.org/abs/2403.07848}{{\tt arXiv:2403.07848}}].

\bibitem{1965On}
J.~Einasto, {\it On the construction of a composite model for the galaxy and on
  the determination of the system of galactic parameters},  {\em Trudy
  Astrofizicheskogo Instituta Alma-Ata} {\bf 5} (1965), no.~12 87--100.

\bibitem{Wang:2019ftp}
J.~Wang, S.~Bose, C.~S. Frenk, L.~Gao, A.~Jenkins, V.~Springel, and S.~D.~M.
  White, {\it {Universal structure of dark matter haloes over a mass range of
  20 orders of magnitude}},  {\em Nature} {\bf 585} (2020), no.~7823 39--42,
  [\href{http://arxiv.org/abs/1911.09720}{{\tt arXiv:1911.09720}}].

\bibitem{Udrescu:2018hvl}
S.~M. Udrescu, A.~A. Dutton, A.~V. Maccio, and T.~Buck, {\it {A deeper look
  into the structure of CDM haloes: correlations between halo parameters from
  Einasto fits}},  {\em Mon. Not. Roy. Astron. Soc.} {\bf 482} (2019), no.~4
  5259--5267, [\href{http://arxiv.org/abs/1811.04955}{{\tt arXiv:1811.04955}}].

\bibitem{Quintana:2022yky}
A.~B. Quintana, F.~J. Llanes-Estrada, and O.~M. Carretero, {\it {Galaxy
  rotation favors prolate dark matter haloes}},  {\em Phys. Rev. D} {\bf 107}
  (2023), no.~8 083524, [\href{http://arxiv.org/abs/2204.06384}{{\tt
  arXiv:2204.06384}}].

\bibitem{Rahaman:2023tkm}
F.~Rahaman, B.~Samanta, N.~Sarkar, B.~Raychaudhuri, and B.~Sen, {\it
  {Traversable wormholes supported by dark matter and monopoles with
  semiclassical effects}},  {\em Eur. Phys. J. C} {\bf 83} (2023), no.~5 395.

\bibitem{Matos:2003nb}
T.~Matos and D.~Nunez, {\it {The general relativistic geometry of the Navarro -
  Frenk - White model}},  {\em Rev. Mex. Fis.} {\bf 51} (2005) 71--75,
  [\href{http://arxiv.org/abs/astro-ph/0303594}{{\tt astro-ph/0303594}}].

\bibitem{Lobo:2020ffi}
F.~S.~N. Lobo, M.~E. Rodrigues, M.~V. de~Sousa~Silva, A.~Simpson, and
  M.~Visser, {\it {Novel black-bounce spacetimes: wormholes, regularity, energy
  conditions, and causal structure}},  {\em Phys. Rev. D} {\bf 103} (2021),
  no.~8 084052, [\href{http://arxiv.org/abs/2009.12057}{{\tt
  arXiv:2009.12057}}].

\bibitem{Tsukamoto:2020bjm}
N.~Tsukamoto, {\it {Gravitational lensing in the Simpson-Visser black-bounce
  spacetime in a strong deflection limit}},  {\em Phys. Rev. D} {\bf 103}
  (2021), no.~2 024033, [\href{http://arxiv.org/abs/2011.03932}{{\tt
  arXiv:2011.03932}}].

\bibitem{Ou:2021efv}
M.-Y. Ou, M.-Y. Lai, and H.~Huang, {\it {Echoes from asymmetric wormholes and
  black bounce}},  {\em Eur. Phys. J. C} {\bf 82} (2022), no.~5 452,
  [\href{http://arxiv.org/abs/2111.13890}{{\tt arXiv:2111.13890}}].

\bibitem{Guo:2021wid}
Y.~Guo and Y.-G. Miao, {\it {Charged black-bounce spacetimes: Photon rings,
  shadows and observational appearances}},  {\em Nucl. Phys. B} {\bf 983}
  (2022) 115938, [\href{http://arxiv.org/abs/2112.01747}{{\tt
  arXiv:2112.01747}}].

\bibitem{Xu:2021lff}
Z.~Xu and M.~Tang, {\it {Rotating spacetime: black-bounces and quantum deformed
  black hole}},  {\em Eur. Phys. J. C} {\bf 81} (2021), no.~10 863,
  [\href{http://arxiv.org/abs/2109.13813}{{\tt arXiv:2109.13813}}].

\bibitem{Ghosh:2022mka}
S.~Ghosh and A.~Bhattacharyya, {\it {Analytical study of gravitational lensing
  in Kerr-Newman black-bounce spacetime}},  {\em JCAP} {\bf 11} (2022) 006,
  [\href{http://arxiv.org/abs/2206.09954}{{\tt arXiv:2206.09954}}].

\bibitem{AbhishekChowdhuri:2023ekr}
A.~Chowdhuri, S.~Ghosh, and A.~Bhattacharyya, {\it {A review on analytical
  studies in Gravitational Lensing}},  {\em Front. Phys.} {\bf 11} (2023)
  1113909, [\href{http://arxiv.org/abs/2303.02069}{{\tt arXiv:2303.02069}}].

\bibitem{Regge:1957td}
T.~Regge and J.~A. Wheeler, {\it {Stability of a Schwarzschild singularity}},
  {\em Phys. Rev.} {\bf 108} (1957) 1063--1069.

\bibitem{Ulhoa:2013fca}
S.~C. Ulhoa, {\it {On Quasinormal Modes for Gravitational Perturbations of
  Bardeen Black Hole}},  {\em Braz. J. Phys.} {\bf 44} (2014) 380--384,
  [\href{http://arxiv.org/abs/1303.3143}{{\tt arXiv:1303.3143}}].

\bibitem{Cruz:2015bcj}
M.~B. Cruz, C.~A.~S. Silva, and F.~A. Brito, {\it {Gravitational axial
  perturbations and quasinormal modes of loop quantum black holes}},  {\em Eur.
  Phys. J. C} {\bf 79} (2019), no.~2 157,
  [\href{http://arxiv.org/abs/1511.08263}{{\tt arXiv:1511.08263}}].

\bibitem{Chowdhury:2020rfj}
A.~Chowdhury and N.~Banerjee, {\it {Echoes from a singularity}},  {\em Phys.
  Rev. D} {\bf 102} (2020), no.~12 124051,
  [\href{http://arxiv.org/abs/2006.16522}{{\tt arXiv:2006.16522}}].

\bibitem{Kobayashi:2012kh}
T.~Kobayashi, H.~Motohashi, and T.~Suyama, {\it {Black hole perturbation in the
  most general scalar-tensor theory with second-order field equations I: the
  odd-parity sector}},  {\em Phys. Rev. D} {\bf 85} (2012) 084025,
  [\href{http://arxiv.org/abs/1202.4893}{{\tt arXiv:1202.4893}}]. [Erratum:
  Phys.Rev.D 96, 109903 (2017)].

\bibitem{PhysRevD.63.084014}
R.~Moderski and M.~Rogatko, {\it Late-time evolution of a charged massless
  scalar field in the spacetime of a dilaton black hole},  {\em Phys. Rev. D}
  {\bf 63} (Mar, 2001) 084014.

\bibitem{PhysRevD.64.044024}
R.~Moderski and M.~Rogatko, {\it Late-time evolution of a self-interacting
  scalar field in the spacetime of a dilaton black hole},  {\em Phys. Rev. D}
  {\bf 64} (Jul, 2001) 044024.

\bibitem{PhysRevD.72.044027}
R.~Moderski and M.~Rogatko, {\it Evolution of a self-interacting scalar field
  in the spacetime of a higher dimensional black hole},  {\em Phys. Rev. D}
  {\bf 72} (Aug, 2005) 044027.

\bibitem{Yang:2022ryf}
Y.~Yang, D.~Liu, Z.~Xu, and Z.-W. Long, {\it {Ringing and echoes from black
  bounces surrounded by the string cloud}},  {\em Eur. Phys. J. C} {\bf 83}
  (2023), no.~3 217, [\href{http://arxiv.org/abs/2210.12641}{{\tt
  arXiv:2210.12641}}].

\bibitem{RevModPhys.83.793}
R.~A. Konoplya and A.~Zhidenko, {\it Quasinormal modes of black holes: From
  astrophysics to string theory},  {\em Rev. Mod. Phys.} {\bf 83} (Jul, 2011)
  793--836.

\bibitem{PhysRevD.75.124017}
E.~Berti, V.~Cardoso, J.~A. Gonz\'alez, and U.~Sperhake, {\it Mining
  information from binary black hole mergers: A comparison of estimation
  methods for complex exponentials in noise},  {\em Phys. Rev. D} {\bf 75}
  (Jun, 2007) 124017.

\bibitem{PhysRevD.35.3632}
S.~Iyer, {\it Black-hole normal modes: A wkb approach. ii. schwarzschild black
  holes},  {\em Phys. Rev. D} {\bf 35} (Jun, 1987) 3632--3636.

\bibitem{PhysRevD.68.024018}
R.~A. Konoplya, {\it Quasinormal behavior of the $d$-dimensional schwarzschild
  black hole and the higher order wkb approach},  {\em Phys. Rev. D} {\bf 68}
  (Jul, 2003) 024018.

\bibitem{PhysRevD.68.124017}
R.~A. Konoplya, {\it Gravitational quasinormal radiation of higher-dimensional
  black holes},  {\em Phys. Rev. D} {\bf 68} (Dec, 2003) 124017.

\bibitem{Berti:2005ys}
E.~Berti, V.~Cardoso, and C.~M. Will, {\it {On gravitational-wave spectroscopy
  of massive black holes with the space interferometer LISA}},  {\em Phys. Rev.
  D} {\bf 73} (2006) 064030, [\href{http://arxiv.org/abs/gr-qc/0512160}{{\tt
  gr-qc/0512160}}].

\end{thebibliography}\endgroup

\end{document}